\documentclass{aa}  

\usepackage{graphicx}
\usepackage{txfonts}
\usepackage{amsmath}
\usepackage{longtable}
\usepackage{longtable,lscape}
\usepackage{lscape}
\usepackage{epsfig}
\usepackage{color}

\begin{document}

   \title{Time-resolved photometry of the young dipper RX~J1604.3-2130A: }

   \subtitle{Unveiling the structure and mass transport through the innermost disk\thanks{Based on observations made with the REM Telescope, INAF Chile, Program ID 37902.},\thanks{Tables A.1 and A.2 are only available in electronic form at the CDS via anonymous ftp to cdsarc.u-strasbg.fr (130.79.128.5)
or via http://cdsweb.u-strasbg.fr/cgi-bin/qcat?J/A+A/}}

   \author{A. Sicilia-Aguilar\inst{1},  C. F. Manara\inst{2}, J. de Boer\inst{3}, M. Benisty\inst{4,5},  P. Pinilla\inst{6}, J. Bouvier\inst{4} }

   \institute{\inst{1} SUPA, School of Science and Engineering, University of Dundee, Nethergate, DD1 4HN, Dundee, UK\\
              \email{asiciliaaguilar@dundee.ac.uk}\\
     \inst{2} European Southern Observatory, Karl-Schwarzschild-Strasse 2, 85748 Garching bei M\"{u}nchen, Germany\\
  \inst{3} Leiden  Observatory,  Leiden  University,  PO  Box  9513,  2300  RA, Leiden, The Netherlands\\
      	        \inst{4} Universit\'{e} Grenoble Alpes, CNRS, IPAG, 38000 Grenoble, France\\
	\inst{5} Unidad Mixta Internacional Franco-Chilena de Astronom\'{i}a (CNRS, UMI 3386), Departamento de Astronom\'{i}a, Universidad de
Chile, Camino El Observatorio 1515, Las Condes, Santiago, Chile\\
          \inst{6} Max-Planck-Institut f\"{u}r Astronomie, K\"{o}nigstuhl 17, 69117, Heidelberg, Germany\\
	        }

   \date{Received August 24, 2019; accepted November 11, 2019}

 
  \abstract
   {RX~J1604.3-2130A is a young, dipper-type, variable star in the Upper Scorpius association, 
suspected to have an inclined inner disk with respect to its face-on outer disk.}
   {We study the eclipses to constrain the inner disk properties.}
   {We use time-resolved photometry from the Rapid Eye Mount telescope and Kepler~2 data to study the multi-wavelength
variability, and archival optical and IR data to track accretion, rotation, and changes in disk structure.}
   {The observations reveal details of the structure and matter transport through the inner disk. The eclipses show 5d quasi-periodicity, with the phase drifting in time and some
periods showing increased/decreased eclipse depth and frequency. 
Dips are consistent with extinction by slightly processed dust grains in an inclined, irregularly-shaped inner disk locked to the star through two relatively stable accretion structures. The grains are located near the dust sublimation radius ($\sim$0.06 au) at the corotation radius, and can explain the shadows observed in the outer disk. The total mass  (gas and dust) required to produce the eclipses and shadows is a few \% of a Ceres mass.  Such amount of mass is accreted/replenished by accretion in  days to weeks, which explains the variability from period to period. Spitzer and WISE variability reveal variations in the dust content in the innermost disk on a few years timescale, which is consistent with small imbalances (compared to the stellar accretion rate) in the matter transport from the outer to the inner disk.
A decrease in the accretion rate is observed at the times of less eclipsing variability and low mid-IR fluxes, confirming this picture. The v$sini$=16~km/s confirms that the star cannot be aligned with the outer disk, but is likely close to equator-on and to be aligned with the inner disk. This anomalous orientation is a challenge for standard theories of protoplanetary disk formation.}
   {}

   \keywords{Stars: individual: 2MASS J16042165-2130284, EPIC 204638512, RX~J1604.3-2130A -- Stars: variables: T Tauri, HAe/Be -- Protoplanetary disks -- Stars: formation 
               }
\authorrunning{Sicilia-Aguilar et al.}

\titlerunning{Eclipses and unstable accretion in RX~J1604.3-2130}

\maketitle
%

\section{Introduction}

Dippers, also called AA Tau-type stars, are young stars with lightcurves characterized by aperiodic dimmings or eclipsing events consistent with variable extinction by circumstellar material \citep{bouvier99}.
Dippers are particularly interesting sources because the occultations of the star by the disk material 
offer an opportunity to explore the structure
and composition of the innermost disk \citep{schneider18} that are inaccessible for most other sources.

RX J1604.3-2130A\footnote{Also known as 2MASS J16042165-2130284, EPIC 204638512.} is a 
solar-type star in the 5-11 Myr old Upper Scorpius Association \citep{preibisch99,pecaut12,pecaut16}
that has been identified as a dipper \citep{ansdell16}. 
It possesses one of the brightest disks detected in the region, which
also has a large inner cavity \citep{carpenter06,carpenter09,dahm09,mathews12,carpenter14,zhang14,dong17}, 
being considered as a transition disk.  The resolved outer disk is nearly face-on, with an estimated
inclination of 6 deg \citep{zhang14,dong17}. The disk contains a substantial
amount of gas \citep{mathews13}, although the cavity shows significant CO depletion \citep{dong17,mayama18}. 

The disk gap could be the result of planetary formation or of the presence of a 
yet-undetected stellar companion. A low-mass (M2) companion, which
is itself a binary with a stellar companion
at 0.082 arcsec \citep{koehler00},  is found at 16" (RX~J1604.3-2130B). Considering the Gaia DR2 parallax 
\citep[6.662$\pm$0.057 arcsec;][]{gaiamission16,gaiadr218} available through VizieR \citep{gaiavizier18},
its distance is 150$\pm$1 pc, which implies that the potential companions are located at about 2400 au and 
12 au (from B), respectively.
Nearby companions of RX~J1604.3-2130A at $>$22 au have been excluded down to 2-3 M$_J$ \citep{kraus08,canovas17},
but there is no record of objects in the innermost region of the gap.

The disk was observed with VLT/SPHERE, revealing a close-to-face-on ring in scattered light about
65\,au in radius \citep{pinilla15,pinilla18}. Dark dents on the scattered light image of the disk rim suggested shadows cast by a highly misaligned inner disk \citep[as it has been observed in other systems;][]{marino15,benisty17,benisty18}. The origin of the shadows in an inclined inner disk has been recently confirmed 
by multi-epoch scattered light observations between 2016-2017 that show that the position angle of the dips varies only slightly \citep[with PA$\sim$83.7$\pm$13.7$^o$ and PA$\sim$265.9$\pm$13.0$^o$, respectively][]{pinilla18}, although the morphology of the shadows is quite variable on timescales of days. 
The presence of a misaligned disk has been also suggested from ALMA gas observations \citep{mayama18} and are also in good agreement with the eclipsing activity observed, but a classical, smooth inner disk cannot explain the
optical and scattered light variability observed. 

In this paper, we use ground-based optical and near-IR photometry from REM/La Silla, together with K2 data and archival photometry and spectroscopy of RX~J1604.3-2130A, to explore the causes behind the 
observed eclipses, and to constrain the structure of the innermost disk and its 
variability timescales. Observations are presented in Section \ref{obs}.
The periodicity of the lightcurves and variability causes in terms of eclipses, accretion, rotation, and variations in the inner disk structure are analyzed in Section \ref{analysis}. The discussion and conclusions are presented in Sections \ref{discussion} and \ref{conclu}.

\begin{figure*}
\centering
\includegraphics[width=0.7\linewidth]{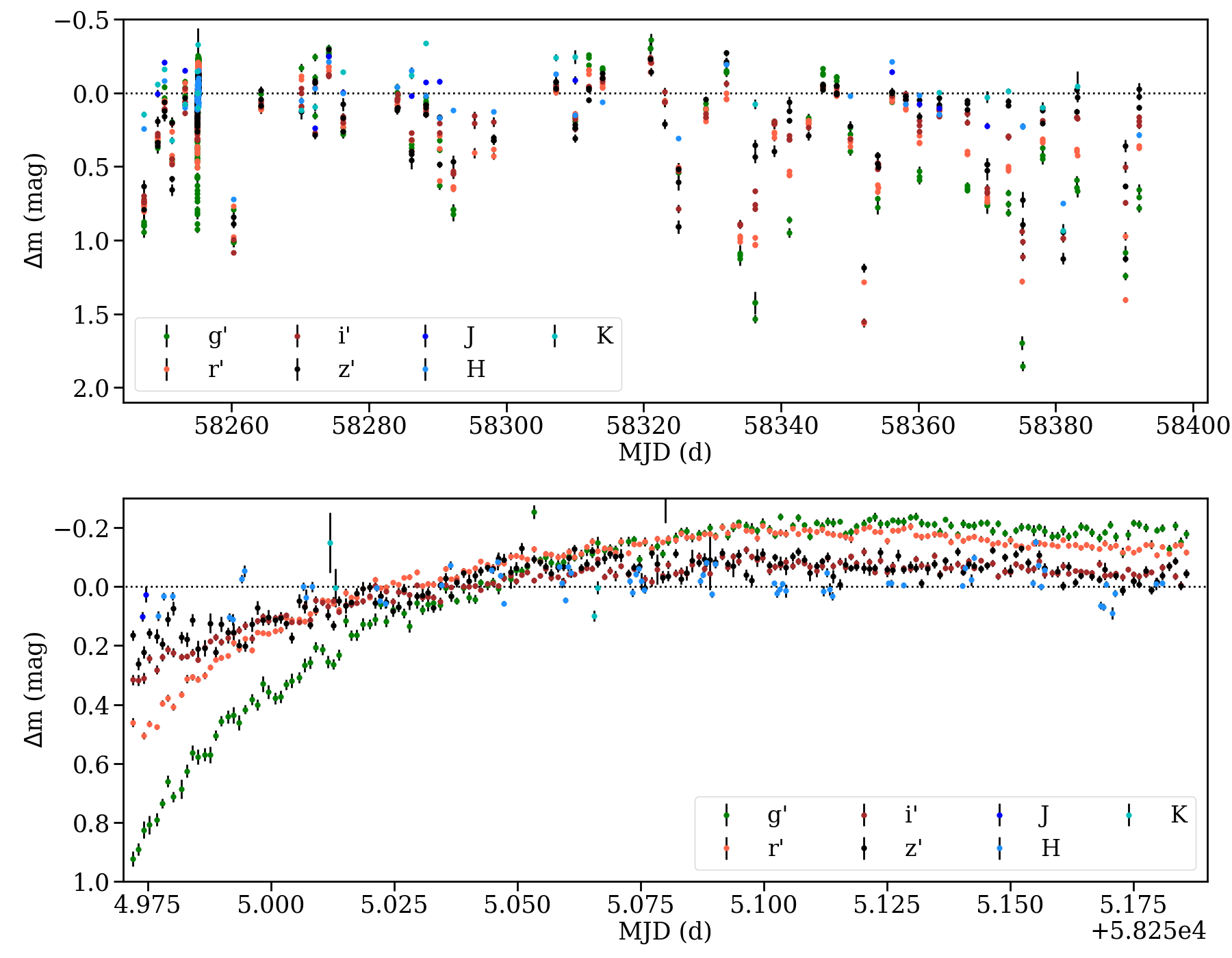}
\caption{REM lightcurve for RX~J1604.3-2130A. The upper panel shows the full lightcurve observed, 
while the lower panel is a zoom in the high-cadence data. All the magnitudes shown are 
relative magnitudes evaluated with respect to the $\sigma$-clipped average magnitude of all datapoints in each band, 
marked as zero point with a dotted line. \label{grizlightcurve-fig}}
\end{figure*}

\section{Observations and data reduction \label{obs}}

\subsection{REM g'r'i'z' JHK observations}

\begin{table*}
\caption{\label{griz-table} Example of REM optical data. All magnitudes are given relative to
those of 58346.021046.}                 
\centering                                     
\begin{tabular}{ccccc}      
\hline \hline                        
MJD & $\Delta$g' & $\Delta$r' & $\Delta$i' & $\Delta$z' \\
(d) & (mag) & (mag) & (mag) & (mag) \\
\hline                             
58247.213152 & 1.031$\pm$0.051 & 0.782$\pm$0.015 & 0.774$\pm$0.025 & 0.658$\pm$0.044  \\
58247.217176 & 1.076$\pm$0.038 & 0.794$\pm$0.066 & 0.790$\pm$0.025 & 0.813$\pm$0.065  \\
58249.213074 & 0.505$\pm$0.037 & 0.340$\pm$0.019 & 0.343$\pm$0.021 & 0.218$\pm$0.034  \\
58249.214534 & 0.457$\pm$0.023 & 0.328$\pm$0.016 & 0.398$\pm$0.018 & 0.361$\pm$0.035  \\
58249.215942 & 0.464$\pm$0.028 & 0.330$\pm$0.015 & 0.327$\pm$0.016 & 0.382$\pm$0.036  \\
58250.226890 & 0.184$\pm$0.019 & 0.080$\pm$0.010 & 0.168$\pm$0.011 & 0.145$\pm$0.016  \\
58250.228301 & 0.164$\pm$0.017 & 0.089$\pm$0.008 & 0.159$\pm$0.013 & 0.084$\pm$0.030  \\
58250.229714 & 0.091$\pm$0.017 & 0.077$\pm$0.013 & 0.165$\pm$0.013 & 0.182$\pm$0.034  \\
58251.278677 & 0.599$\pm$0.020 & 0.447$\pm$0.013 & 0.499$\pm$0.011 & 0.606$\pm$0.021  \\
58251.280120 & 0.579$\pm$0.024 & 0.500$\pm$0.024 & 0.536$\pm$0.012 & 0.681$\pm$0.039  \\
\hline
\end{tabular}
\tablefoot{The complete table is given in the online appendix (Table \ref{griz-app-table}).}
\end{table*}

RX~J1604.3-2130A was observed with the Rapid Eye Mount (REM) 60cm telescope\footnote{http://www.rem.inaf.it/} in La Silla, 
as part of a DDT proposal. With its two instruments, ROS2 and REMIR, REM 
can obtain nearly-simultaneous images in the Sloan  g',r',i',z' and the IR JHK filters over a 10x10 arcmin$^2$ field. The observations took place at irregular intervals during approximately 5 months, from  2018 May 09 to 2018 October 01 (MJD 58247.21 - 58392.08). The observations were repeated at intervals between half an hour and few days, with a period of high-cadence on 58255.09 (2018 May 16-17) during which the images were taken about every 2 min. For the optical bands, a total of 338 images were obtained, among which 168 of them belong to the high-cadence dataset. The total exposure time for the optical images was 5s. Data reduction was performed with the automated REM pipeline, and the images were aligned using Astrometry.net.
The IR observations were obtained during the same epochs, but due to technical issues,
only 42 datasets were completed for J, 43 for K, and 142 for H. Typically, a total of 5 dithered 3s exposures were taken in each case, although some of them have less dithers if the imaging failed (especially in bad nights). Sky images were also acquired, although some of the final images have less dithers due to lack of quality of some of them. The images were reduced, combined, sky-subtracted and aligned using the automated REM pipeline, and aligned using Astrometry.net.

Aperture photometry was performed using
$iraf$ task $noao.digiphot.apphot$. For the optical data, a relative calibration was 
performed for each filter by 
comparing all observations to the data taken on MJD 58346.021 (one of 
the best nights, based on seeing/FWHM).  An iterative process based on the 
median and standard deviation of the magnitude difference 
\citep{sicilia08,sicilia17} was used on all
stars in the field to identify the non-variable comparison stars. 
The main issue is 
that there are few comparison stars in the field, 
and all of them are fainter than the target star. This is due to 
RX~J1604.3-2130A  brightness (Gaia G 11.87 mag) requiring low exposure
times to avoid saturation. We thus imposed 
quality limits on the calibration, rejecting all those for which 4 or less comparison 
stars were identified and those where the calibration had magnitude-dependent offsets
or very large errors. Typically, we found between 5 and 15 comparison stars for g' and 
r', and between 10 and 27 comparison stars for i' and z', spanning a range of 2-4 
magnitudes around (but mostly fainter than) the object. The nights for which the 
calibration fails are typically those with poor seeing and poor weather conditions 
that make it hard to detect enough  field stars. This 
results in 205 dates for which all four optical filters are complete. The final 
data (relative magnitudes) are listed in Table \ref{griz-table} with the complete table given in 
Appendix \ref{data-app}. The uncertainties provided in the
tables and figures include the photometric uncertainty and the 
uncertainty in the relative calibration. In general, the uncertainty in the relative calibration
dominates the value in cases with few comparison stars.

The absolute calibration of the g'r'i'z' data was done using griz data from PAN-STARRS\footnote{Note that the field is not included in the Sloan Digital Sky Survey.}  \citep{chambers16,flewelling16,magnier16}. 
The procedure for the absolute calibration was similar to the relative calibration, comparing
the data from MJD 58346.021 with those of PAN-STARRS. The g'r'
calibration is quite robust (2\% and 4\% uncertainties, respectively), but for i' there is a large amount of scatter and for z' only 4 reference stars could be identified. In addition, we find that there are
color terms for all filters except g'. The color terms are particularly large for i' and z'. Since there are
no stars as bright as RX~J1604.3-2130A in the field and very few non-variable stars in any magnitude range,
the absolute calibration is very uncertain and the errors could be  $>$50\% in i' and z'. The magnitudes in the reference night MJD 58346.021 are thus g'=12.47$\pm$0.02\,mag, r'=11.01$\pm$0.04:\,mag, i'=11.7:\,mag, and z'=11.0:\,mag. 
Because the magnitudes in r',i', and z' are estimated from much fainter stars, we label them as uncertain (:) and 
treat them as merely indicative, and we focus the discussion on relative REM magnitudes and relative color variations. In total, there 
are 261 datapoints in g', r' and i, and 270 in r' observations.
                         
For the JHK data, we followed a similar procedure. The data were calibrated
against those of MJD 58363.04 (one of the best nights for which all three IR filters were obtained), which was also calibrated against 2MASS data \citep{skrutskie06}. The 
REM and 2MASS JH filters are essentially identical, while the relation between the
standard Johnson K filter and the 2MASS K$_s$ is K=K$_s$+0.044\,mag \citep{bessell05}. The 
systematic errors in the absolute calibration are 4\% for J and H and 2\% for K. There are no evident color terms
between the 2MASS and REM filters, but
because the data quality of the JHK images is in general worse than in the optical, 
there are fewer stars for comparison. Following the same quality criteria, we only
have complete JHK data on 10 epochs, although the H lightcurve is much more complete 
(73 points) while J and K have only 8. The final results are listed in Table \ref{JHK-table}
(complete table in Appendix \ref{data-app}).

\begin{table}
\caption{\label{JHK-table} Example of JHK REM observations. }                 
\centering                                     
\begin{tabular}{ccccc}      
\hline \hline                        
MJD & J & H & K \\
(d) & (mag) & (mag) & (mag)  \\
\hline                             
58250.227 & 8.920$\pm$0.020 & 8.205$\pm$0.007 & 7.895$\pm$0.012  \\
58253.213 & 8.974$\pm$0.016 & 8.388$\pm$0.015 & 8.133$\pm$0.029  \\
58272.150 & 9.365$\pm$0.014 & 8.256$\pm$0.041 & 8.152$\pm$0.026  \\
58276.156 & 9.122$\pm$0.022 & 8.287$\pm$0.007 & 7.916$\pm$0.011  \\
58286.128 & 9.145$\pm$0.018 & 8.135$\pm$0.023 & 7.938$\pm$0.024  \\
58288.245 & 9.055$\pm$0.013 & 8.307$\pm$0.025 & 7.720$\pm$0.009  \\
58309.989 & 9.042$\pm$0.028 & 8.438$\pm$0.009 & 7.812$\pm$0.047  \\
58363.042 & 9.229$\pm$0.012 & 8.431$\pm$0.007 & 8.054$\pm$0.009  \\
\hline
\end{tabular}
\tablefoot{The complete table, including nights with only partial data (one or two filters) is given in the online appendix (Table \ref{JHK-app-table}). Here we only list the datapoints for which all three JHK magnitudes are available. Note that because the JHK exposures are not fully simultaneous, the MJD indicated is the one of the J-band observation. 
The data are calibrated using 2MASS. The uncertainties shown are those resulting from the relative calibration, and
 do not contain an extra 2-4\% uncertainty due to the absolute calibration (see text).}
\end{table}

The final REM ligthcurve is displayed in Figure \ref{grizlightcurve-fig}. 
The optical data show the typical dimming events described in \citet{ansdell16}.
The JHK data follows the behavior observed in the optical, although they are more scarce. The high-cadence data reveals the star emerging from one of the eclipses. We find that the eclipses observed by REM are deeper than previously reported \citep[0.57 mag based on K2 data;][]{ansdell16}. Even though our filters are significantly narrower than the K2 filter, the maximum depth varies between 0.4--1.8 mag in g', 0.3--1.5 mag in r', 0.3--1.5 mag in i', and 0.2--1.2 mag in z'. The variations in H (the only IR filter for which we have enough eclipse data) are up to 0.2-0.7 mag. There are also smaller variability events, but we detect at least 10 deep eclipses during our observations (all of which are recovered in multiple bands), in addition to other shallower ones similar to those reported by \citet{ansdell16}.  The decrease in depth with increasing wavelength suggests extinction events, which we will explore in Section \ref{analysis}.

\subsection{Other optical lightcurves: K2 and CSS}

\begin{figure*}
\centering
\begin{tabular}{c}
\includegraphics[width=0.9\linewidth]{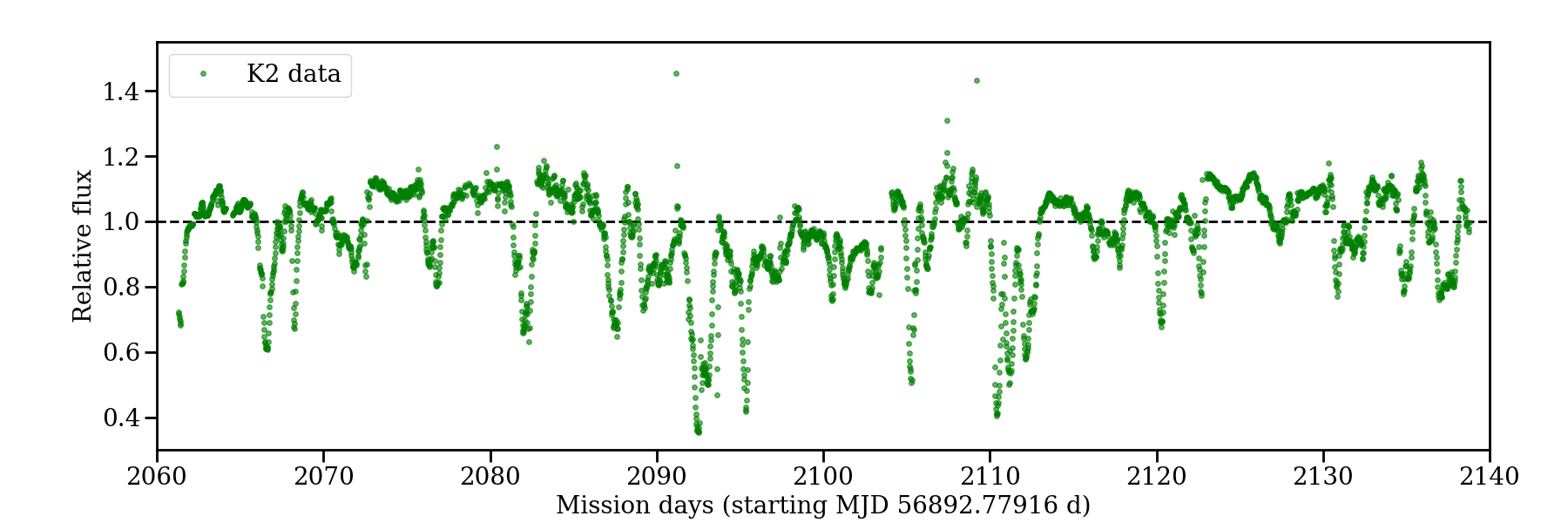} \\
\includegraphics[width=0.9\linewidth]{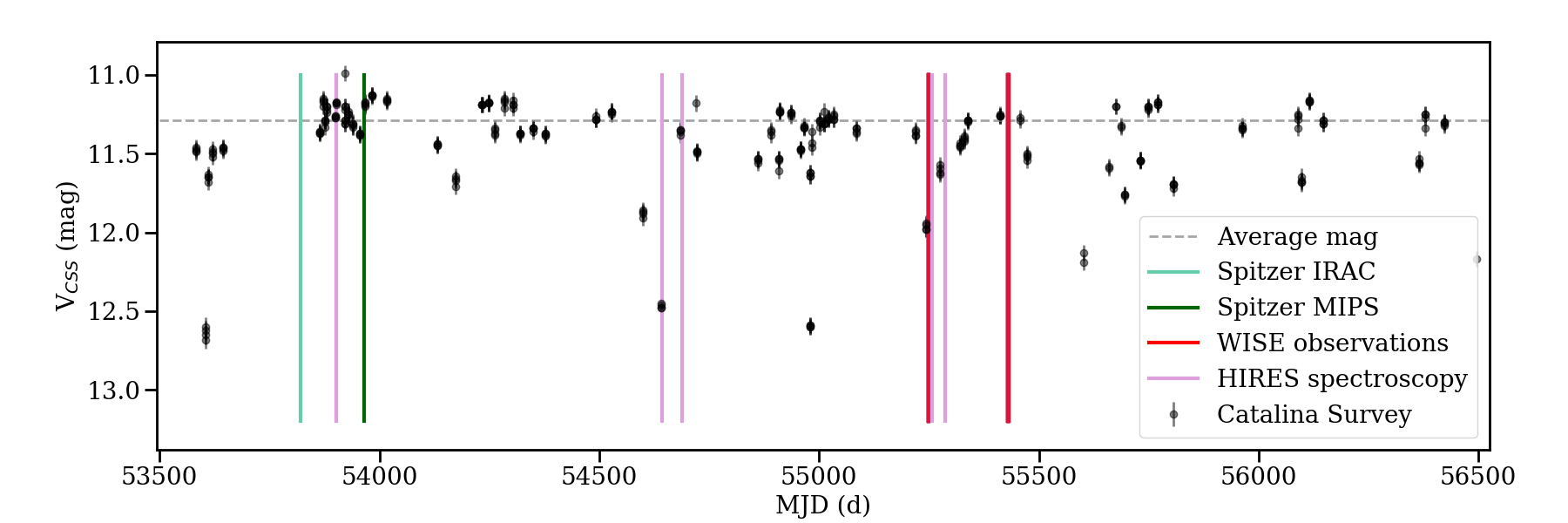} 
\end{tabular}
\caption{Top: K2 lightcurve for RX~J1604.3-2130. The typical uncertainties are smaller than the dots. Bottom: Catalina Survey DR2 lightcurve. Since the Catalina data covers the epochs of the Spitzer, WISE and HIRES spectroscopy 
observations, we have marked them in the figure as
vertical lines (see text).  In both cases, the average flux (magnitude) is shown as a dotted line.\label{K2lightcurve-fig}}
\end{figure*}

RX~J1604.3-2130A was observed by K2 as part of the Ecliptic Plane Input Catalog \citep[EPIC;][]{huber16}
as source EPIC 204638512. The K2 data was obtained from 
the Mikulski Archive for Space Telescopes (MAST\footnote{https://archive.stsci.edu/missions/k2/lightcurves/c2/204600000/ 38000/ktwo204638512-c02\_llc.fits}) at the Space 
Telescope Science Institute\footnote{The full data used in this work can be accessed 
at https://doi.org/10.17909/t9-ap7q-e405}. Although some K2 data are affected by interlopers, for RX~J1604.3-2130A the MAST K2SFF public lightcurves have been validated so that
there is no need of further corrections \citep{ansdell16}.
The data were acquired between 2014 August 23 - 2014 November 10 (MJD 56892.78 - 
56971.55, thus about 4 years before the REM data. The K2 data are essentially uniformly 
sampled, with a sampling period of 30 min, and are presented as relative fluxes, thus having values around 1 out of eclipse. Figure \ref{K2lightcurve-fig} displays the complete K2 lightcurve, which shows the irregularly-shaped dimming events up to 0.57 mag reported by \citet{ansdell16}. We also note that, besides the dips, there are several cases where a sudden brightness increase is observed. These may be stellar flares and are discussed in Appendix \ref{flares-app}.

RX~J1604.3-2130A has been observed by the Catalina Sky Survey \citep[CSS;][]{drake09}, Data Release 2\footnote{http://nesssi.cacr.caltech.edu/DataRelease/}. The CSS archive contains 293 photometry points distributed over nearly 8 years, 
from 2005-08-01 to 2013-07-22 (MJD  53583.454 - 56495.535; see Figure \ref{K2lightcurve-fig}). The object ID in the survey is SSS\_J160421.7-213028.
The sampling is very sparse compared
to the relevant timescales, but it shows the same behavior detected in the REM and K2 data, with sudden dimmings that in some
cases go down by nearly 1.4 magnitudes in V and some periods of relative stability. Some of the data are very close to the saturation limit (11 mag), so that the highest magnitudes may be uncertain, but the eclipse data are well below saturation.
Although no further information can be obtained from these data regarding periodicity, they essentially confirm the behavior observed
and the fact that the eclipse depths are highly variable and persistent. Note that the Catalina V filter 
has a non-negligible color
term\footnote{http://nesssi.cacr.caltech.edu/DataRelease/FAQ2.html\#reference}. The color term is stronger for very 
red objects, so it is likely affecting the eclipse depth. Considering the color variations observed
with REM and the typical colors for a K3-type star \citep{KH95}, the maximum eclipse depth in V (Cousins system) may be
shallower by up to 0.3-0.4 mag with respect to the value observed in the CSS lightcurve.

\subsection{Archival optical spectroscopy}

With the aim to understand the causes of variability, we also need to constrain rotation and
accretion, which are two of the major causes leading to magnitude fluctuations observed in 
young stars. We thus study  archival 
high-resolution spectroscopy in the analysis. RX~J1604.3-2130A was observed 6 times with the High Resolution Echelle Spectrometer \citep[HIRES;][]{vogt94} between June 2006 and April 2010, 
available through the Keck Observatory Archive (KOA\footnote{ http://koa.ipac.caltech.edu/}). Exposure times, coverage, and resolution varied and are listed in Table \ref{spectraMJD-table}. The data were reduced using the 
automated MAKEE pipeline\footnote{http://www.astro.caltech.edu/~tb/makee/}. The automated reduction includes bias and flat field correction and calibration using
a ThAr lamp. The long-slit data were used to extract and subtract the sky spectrum. The spectra were extracted in vacuum wavelength and subsequently 
transformed using PyAstronomy\footnote{https://github.com/sczesla/PyAstronomy} routine $vactoair2$. No flux calibration was performed.

\begin{table}
\caption{HIRES/Keck spectroscopy summary,  including wavelength coverage and resolution (R). \label{spectraMJD-table}}
\centering
\begin{tabular}{cccc}
\hline \hline
MJD & Exp.Time & Coverage & R\\
(d) & (s) & (\AA)  & \\
\hline
53902.275 & 30 & 4775-9200 & 35800\\
53902.278 & 300 & 4775-9200 & 35800\\
54642.417 & 600 & 4700-9200 & 47700 \\
54689.304 & 500 & 3800-8000 & 47700 \\
55256.556 & 500 & 3800-8000 & 47700 \\
55287.614 & 900 & 4300-8500  & 47700\\
\hline
\end{tabular}
\tablefoot{
Coverage is not continuous, there are gaps between orders.}
\end{table}

Besides looking for variability signatures, the spectrum with the best S/N 
(MJD 55287.614) was also used to confirm the projected rotational velocity (v$sini$) of the object.
We selected the region between 5500-5800$\AA$, which is relatively devoid of
both accretion- and activity-related emission lines and telluric
lines \citep{curcio64} and measured the rotational and radial velocity by 
cross-correlating the object spectrum with 3 different rotational standards 
with similar spectral types\footnote{Taken from: http://obswww.unige.ch/\%7Eudry/std/stdnew.dat} that had been observed with Keck under similar conditions. These included
HD 114386 \citep[K3, also used by][]{dahm12}, HD 10780 (K0), and HD 151541 (K1).

PyAstronomy task $rotBroad$ was used to create artificially broadened templates, and  $crosscorRV$ was used to obtain the cross-correlation. The location of the cross-correlation peak was used to determine the radial velocity (v$_{rad}$), and the width of the cross-correlation function was compared with that of the broadened templates to obtain the rotational velocity v$sini$. We obtained v$_{rad}$=-6.8$\pm$0.1 km/s
and v$sini$=16.2$\pm$0.6 km/s.  Both are in good agreement with \citet{dahm12}, and confirm that the star is a relatively fast rotator compared to young stars with
similar spectral types \citep[e.g.][]{sicilia05,weise10}, especially if we take into account that the system is accreting
\citep{dahm12}. Although the rest of datasets are significantly worse in quality, the
results derived are consistent (see a summary in Appendix \ref{kinematics-app}).

\subsection{Further archival data and stellar parameters}

RX~J1604.3-2130A has been repeatedly observed in the mid-IR by WISE and Spitzer \citep{carpenter06,luhman12,esplin18}. The star has some signs of intriguing mid-IR variability, changing from photospheric colors as observed with Spitzer on MJD 53820, to clear mid-IR excess as observed with WISE after MJD 55249 
\citep{luhman12}. We thus include the Spitzer and WISE data in the discussion, taking the IRAC photometry values
reported by \citet{carpenter06} and the available AllWISE Multiepoch Photometry\footnote{http://wise2.ipac.caltech.edu/docs/release/allwise/} \citep{wright10} lightcurve data provided by IRSA\footnote{https://irsa.ipac.caltech.edu/Missions/wise.html}.

%

To examine the disk structure and behavior, we 
use the multiwavelength data available within VizieR to
construct the spectral energy distribution (SED\footnote{http://vizier.u-strasbg.fr/vizier/sed/}; see Table \ref{sed-table}
for wavelengths, fluxes and references). The available data
includes SDSS, Gaia, POSS-II, Hipparcos, and SkyMapper optical data; 2MASS, POSS-II, UKIDSS, and Vista near-IR
data; and Spitzer, Akari, IRAS and IRAS mid- and far-IR data, together with the 880$\mu$m datapoint from ALMA.
The magnitude data are converted to fluxes using the same calibrations as VizieR \citep{bessell88,fukugita96,cohen03}.

Finally, in the whole discussion we use the most recent estimates of the stellar parameters derived from X-Shooter
spectroscopy (see Manara et al. in prep),
which give a K3 spectral type, T$_{eff}$=4730 K, L$_*$=0.90 L$_\odot$\footnote{ These stellar parameters
are not significantly different from previous estimates, e.g. from \citet{preibisch99}, except for the fact that the star now appears to be more luminous, maybe because of having been previously measured during eclipse.} (so the stellar radius is R$_*$=1.4 R$_\odot$), and M$_*$=1.24 M$_\odot$ \citep[using the spectral type calibration and tracks from][respectively]{luhman03, baraffe15} and estimates an accretion
rate of 3e-11 M$_\odot$/yr at a time when the star was in a relatively bright state.  Note that the accretion rate is on the limit of what can be detected with X-Shooter \citep[the object is classified as a potential accretor by][based on its weak accretion features]{dahm12}, which adds uncertainty to the measured value, although the line profiles seen with
HIRES show clear accretion. Using these stellar parameters, we can also estimate that the dust sublimation
radius (for T=1500-1000 K) is located at about 0.06-0.15 au ($\sim$10-22 R$_*$).

\section{Analysis \label{analysis}}

\begin{figure}
\centering
\begin{tabular}{c}
\includegraphics[width=0.7\linewidth]{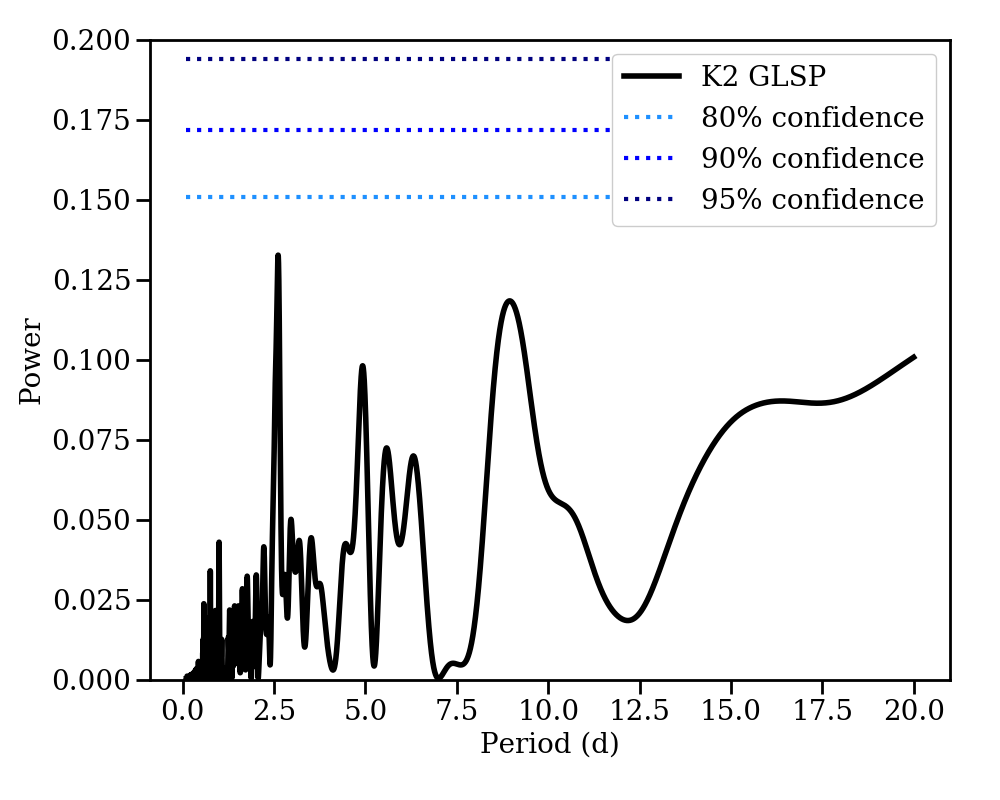}\\
\includegraphics[width=0.7\linewidth]{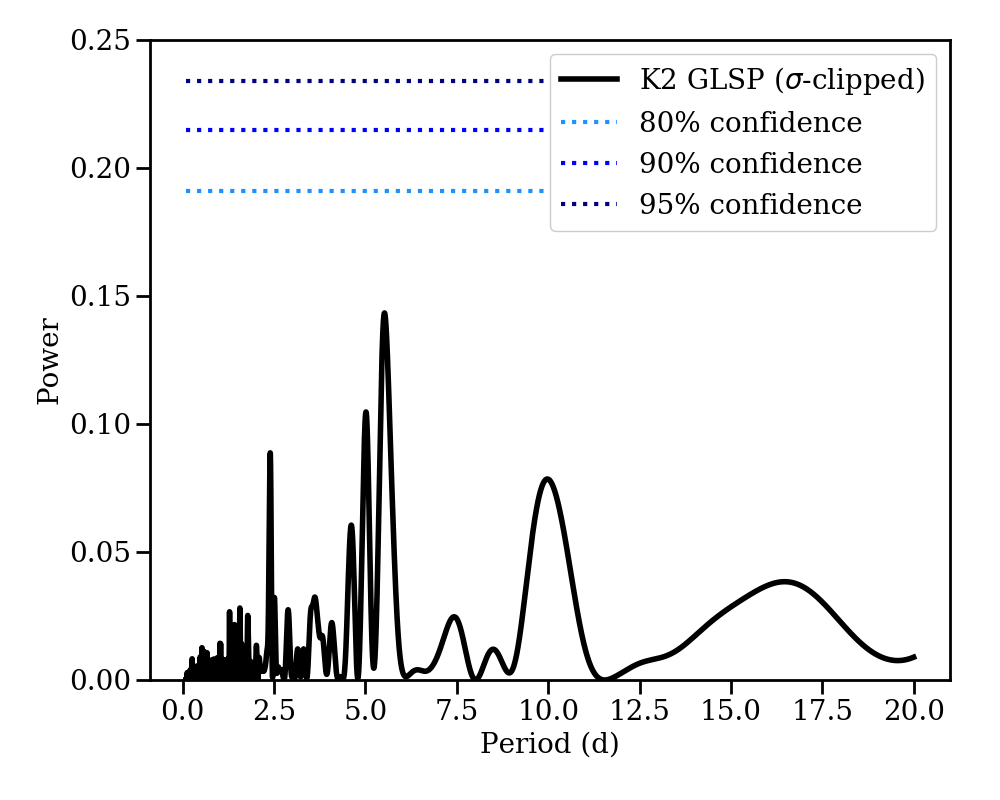}\\
\end{tabular}
\caption{Top: GLSP for the complete K2 dataset. Bottom: GLSP for the $\sigma$-clipped dataset
(out-of-eclipse data). The significance levels are estimated
according to a red-noise model with correlation parameter $\alpha$=0.98 and the same uncertainty distribution and
sampling as observed in the data (see text), so that the observed peaks are not
significant. \label{K2fullGLSP-fig}}
\end{figure}

\begin{figure}
\centering
\includegraphics[width=0.9\linewidth]{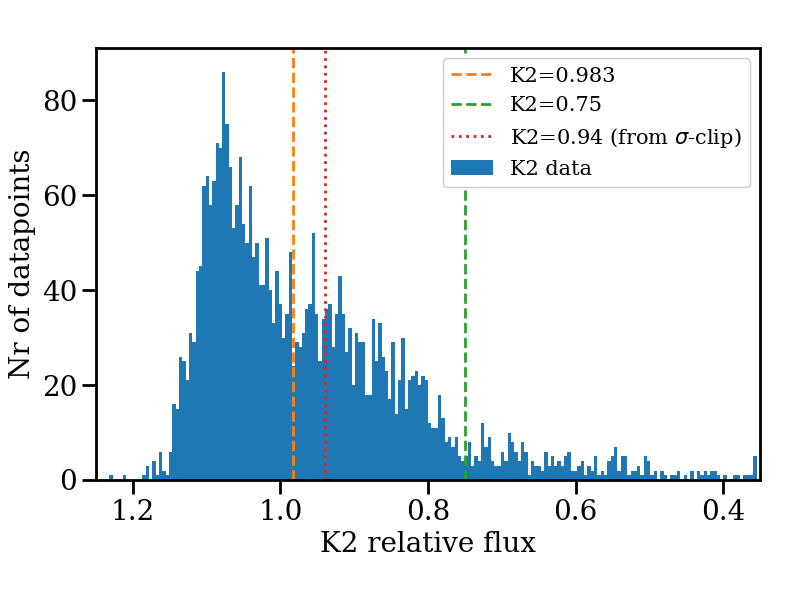}\\
\caption{Histogram of the K2 data by magnitude. Note that there is no clear separation between the "on-eclipse" and
"off-eclipse" parts, except for very deep eclipses (relative flux$<$0.75, green line). The red dotted line marks the
separation from the $\sigma$-clipping filter. Selecting down to a different level (e.g. 0.983, orange line, where the distribution appears to slightly flatten out) does not introduce any significant change.\label{K2histo-fig}}
\end{figure}

\subsection{Periodicity analysis}

A period of about 5d (albeit very uncertain) has been suggested as the 
rotational period of RX~J1604.3-2130A from K2 data \citep{ansdell16,rebull18}.
Here we revisit the periodicity in the K2 lightcurve to examine whether it is 
most likely due to rotation, or related to the obscuration events.
The lightcurve is extremely irregular, suggesting variations in both
the phase, the period, and the amplitude of the modulations and the presence of correlated, non-Gaussian noise.
Therefore, we take two approaches to search for periodical signatures: generalized Lomb-Scargle
periodograms \citep[GLSP;][]{scargle82,horne86,zechmeister09} and wavelet analysis \citep{torrence98,liu07}. 

Simple GLSP fail when applied to quasi-periods, so for the first approach we 
use stacked GLSP \citep[SGLSP;][]{mortier17}, where the data are 
filtered around each single date to study periodicity only over a limited 
number of days. Repeating the exercise over time, changes in the period and phase can be
tracked.
Since the data distribution is highly
non-Gaussian and strongly correlated, a red noise model is needed to assess
the significance of the signatures. The red noise model was derived from the correlation between
consecutive datapoints, parameterized by $\alpha$, the slope of the correlation between one 
datapoint and the next. For K2 data, we find $\alpha$=0.98. 
We then simulated data following a similar distribution where each 
value depends on the previous
one (as given by $\alpha$) plus an stochastic component with values drawn from random numbers 
following the standard deviation of the observed data.
For each K2 (S)GLSP, we computed 1000 red-noise simulations
with the same number of datapoints, the same sampling rate, and the same red-noise model but without any periodic 
signature and used their periodograms to derive the confidence intervals.

\begin{figure*}
\centering
\begin{tabular}{ccc}
\includegraphics[width=0.31\linewidth]{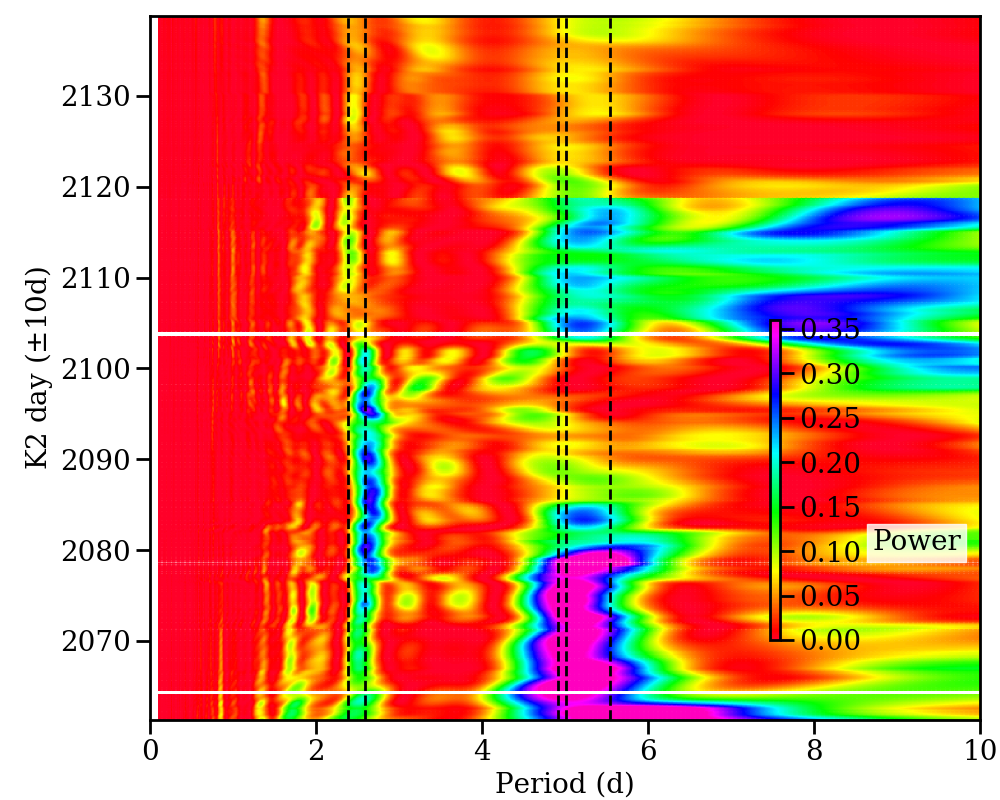} &
\includegraphics[width=0.31\linewidth]{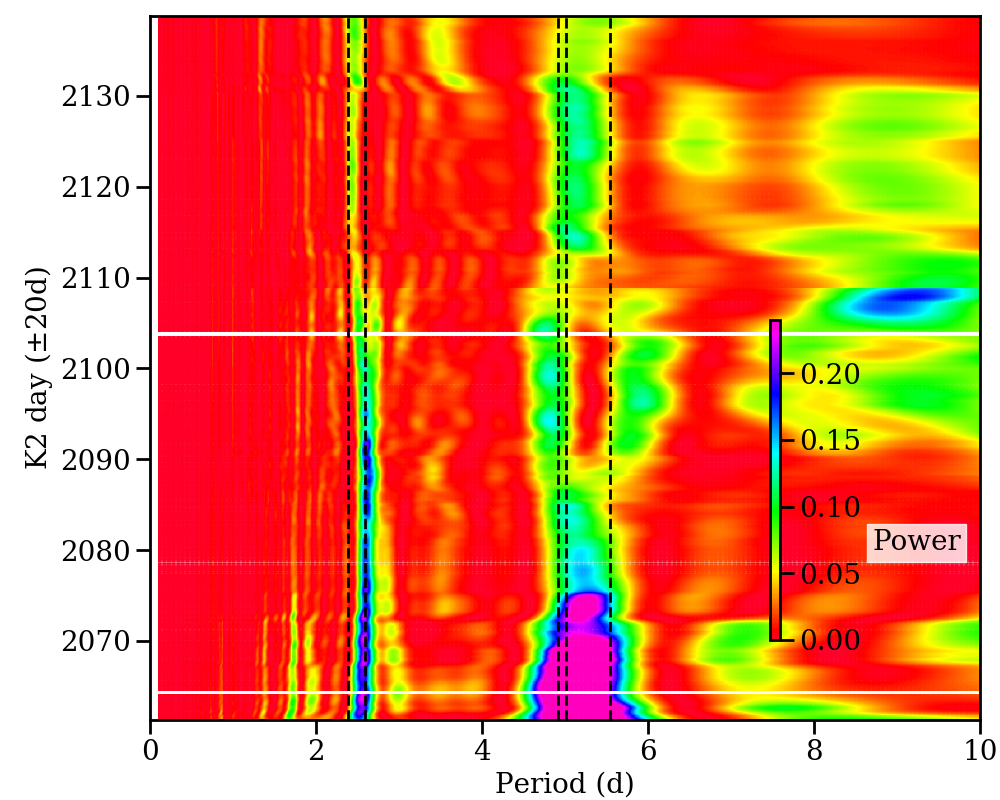} &
\includegraphics[width=0.31\linewidth]{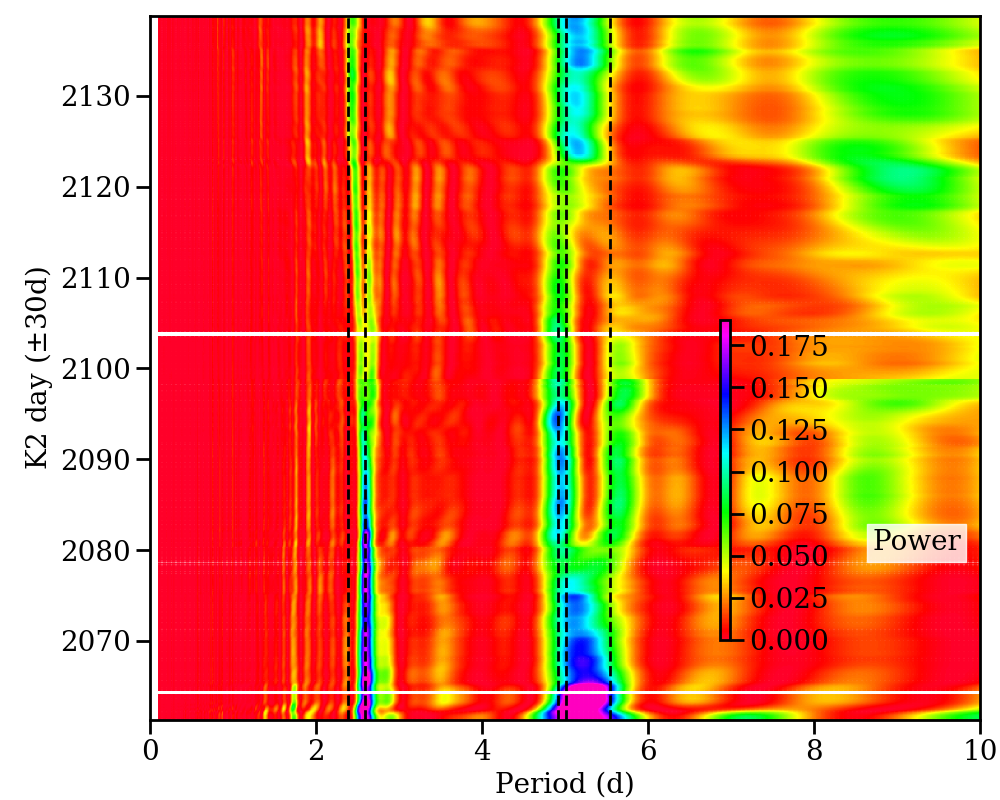} \\
\end{tabular}
\begin{tabular}{cc}
\includegraphics[width=0.31\linewidth]{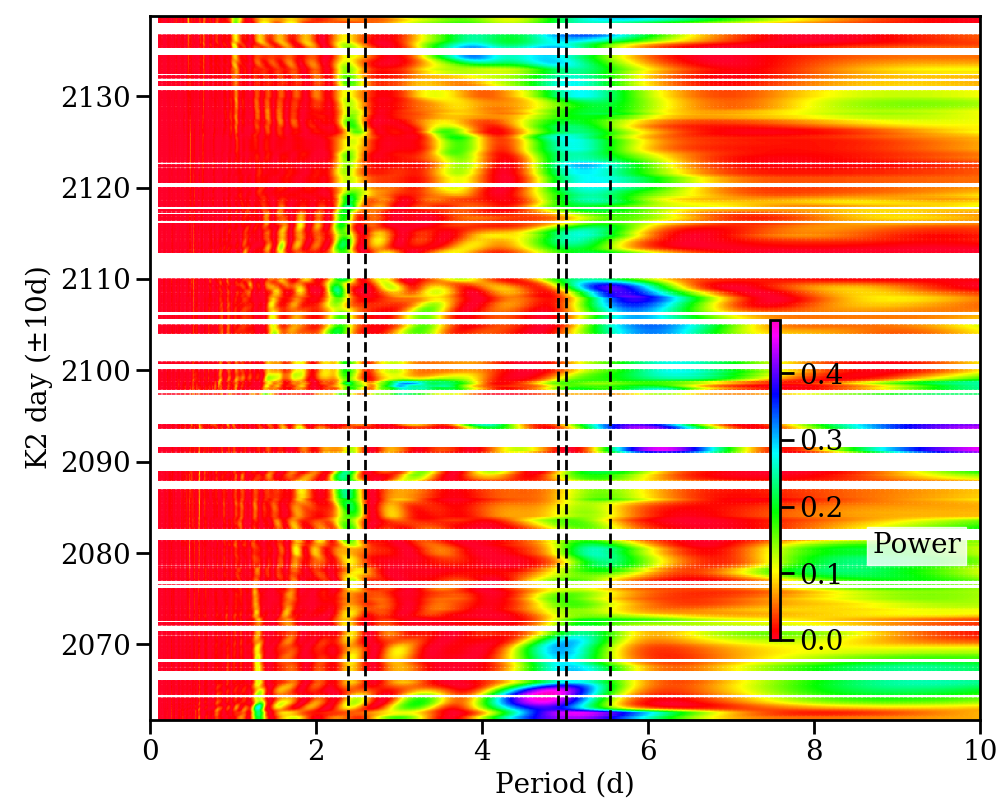} &
\includegraphics[width=0.31\linewidth]{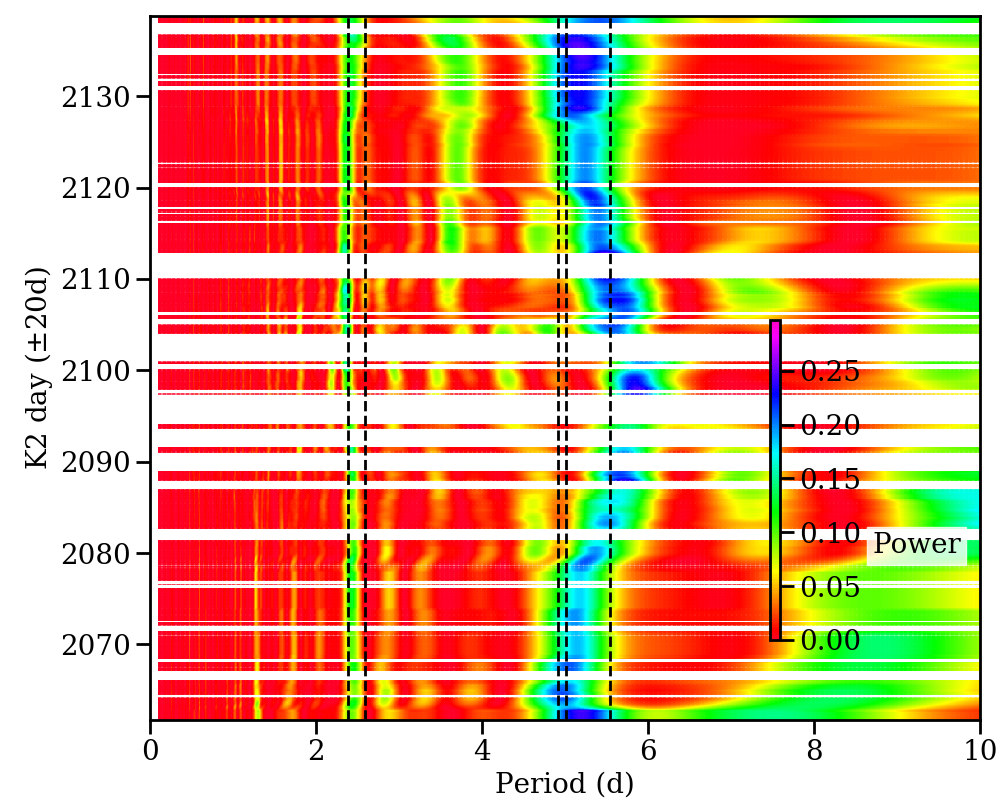} \\
\end{tabular}
\caption{Stacked GLSP for the full K2 data (top three panels, stacking at intervals of $\pm$10d, $\pm$20d, $\pm$30d) and the $\sigma$-clipped data (out-of-eclipse data; bottom two panels, stacking at intervals of $\pm$10d and $\pm$20d). The most significant
periods detected in the individual GLSP are marked with vertical lines (see text for discussion). 
In each panel, the x axis shows the periods and the y axis shows the date (in K2 mission days) around which we
consider the time interval to estimate the SGLSP. Dates for which no data are available are left blank. 
The color scale is set such that purple
is equivalent to 95\% significance and dark blue is equivalent to 90\% significance for a red-noise model with the same number of points, 
similarly distributed (see text).   \label{K2stackedGLSP-fig}}
\end{figure*}

A GLSP including all the available data suggests some signals with
periods 2.5d (which could correspond to half the 5d period), 5d, and 9d, but none of them is 
highly significant (see Figure \ref{K2fullGLSP-fig}). The 5d periodic signature is reported in the literature as a
rotational period, so we first examined the data belonging to the out-of-eclipse
phase and the data belonging to the eclipses separately. Defining the "out-of-eclipse phase" is not easy
in a lightcurve that  does not show a clear 
difference between the "in-eclipse" vs "off-eclipse" parts (see Figure \ref{K2histo-fig}). 
We thus cleaned the data using a $\sigma$-clipping algorithm to calculate the
mean and standard deviation, and then removed all points that are beyond 3$\sigma$ from this value. The analysis
of the off-eclipse data revealed that the
periodic signatures are stronger when the full dataset is considered, and thus the 
(quasi-)periodicity is strongly linked to the eclipses, and not only to rotation.

\begin{figure*}
\centering
\includegraphics[width=0.8\linewidth]{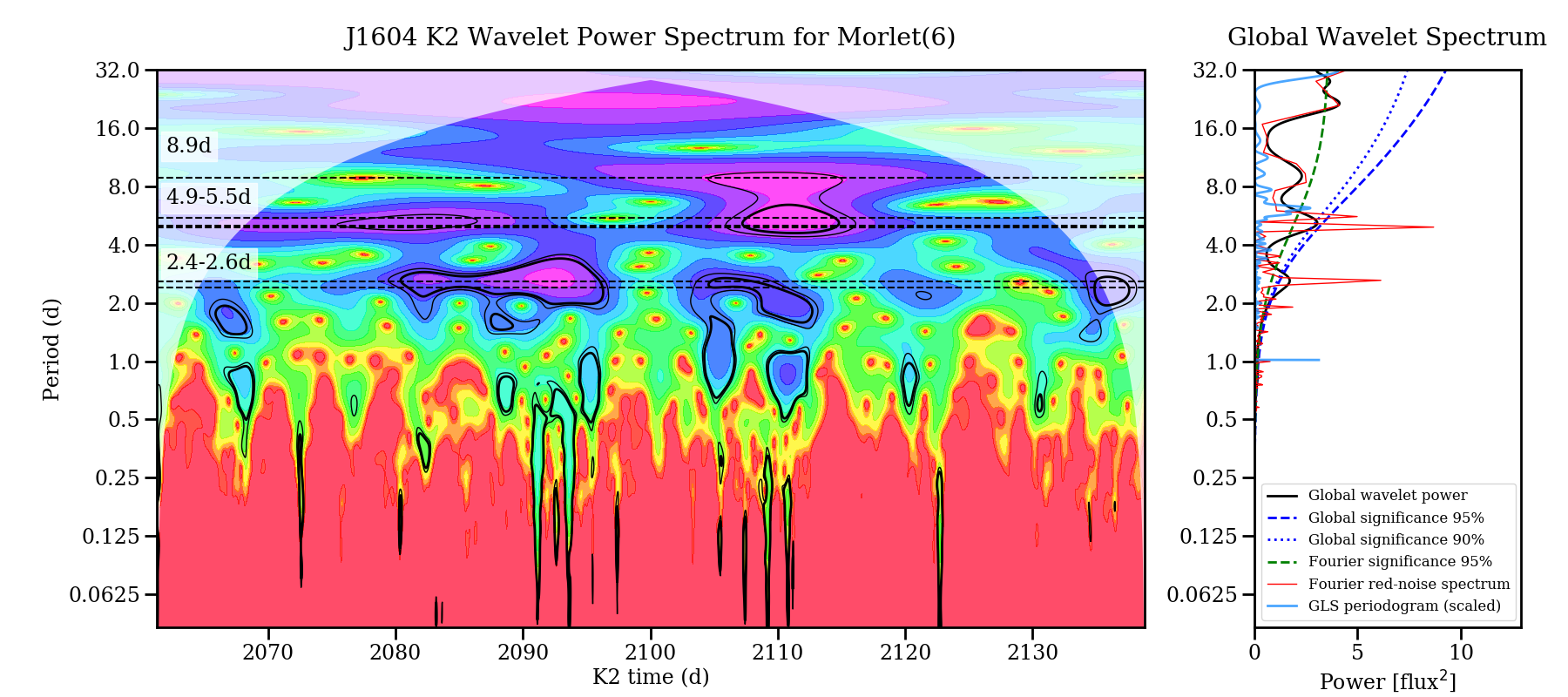}
\caption{Wavelet power spectrum for the K2 data on RX~J1604.3-2130. The left panel shows the wavelet spectrum in time.  The color scale is set between minimum and maximum power, and the global significance limits are marked as thick black (95\% confidence) and thin black (90\% confidence) contours, calculated for a red noise data model with $\alpha$=0.98. The regions where edge effects could be significant are masked out. Horizontal lines mark the positions of the GLSP peaks found with K2 and REM (at 8.9d, 5.5d, 5.0d, 4.9d, 2.59d, 2.39d). The right panel shows the global wavelet power together
with the significance, the noise spectrum, and the scaled GLSP for comparison.\label{wavelet-fig}}
\end{figure*}

We then stacked the data for different numbers of days around each point,
and calculated SGLSP in intervals of $\pm$10d, $\pm$20d, $\pm$30d, and $\pm$40d. The number of
days was selected to be large enough to detect the potential periods observed in the full
collection of data, up to the limit of $\pm$40d that includes essentially the whole dataset
and tends to the full-data GLSP. Figure \ref{K2stackedGLSP-fig} displays the results. 
The data reveals a 5d period in the stacked $\pm$10d and $\pm$20d diagrams,
which progressively dilutes when more data are added. The period is not always present, 
and the peak changes between 4.8-5.5d, which makes it more plausibly a quasi-period
related to a rapidly variable phenomenon
than a typical rotational period, even if both may be connected. The 2.5d period is also present, although it 
has a lower significance except during the time of increased eclipse activity (approximately, from
mission day 2085 to 2105), when eclipses occur at a higher rate. 
The 9d period is very broad and not well-defined.

For the second periodicity estimate, we use wavelet analysis based on 
a Morlet function with $\omega_0$=6, which typically 
offers the best results for complex datasets \citep{torrence98}. The Morlet wavelet $\Psi$($\eta$)  \citep{grinsted04} is very similar to a sinusoidal function tappered by a Gaussian, written as
\begin{equation}
\Psi(\eta) = \frac{e^{i\omega_0 \eta}}{\pi^{1/4}}  e^{-\eta^2/2}.  \label{morlet6-eq}
\end{equation} 
Here, the time dependency is wrapped in the dimensionless parameter $\eta$, which takes values that are
multiple of powers of 2 assigned through the dimensionless time series\footnote{Since the data are equally-spaced, the analysis is done considering their order number rather than the physical time.}.  In essence, 
the SGLSP and wavelet analysis are quite similar, with
the main difference being that SGLSP uses a square passband to filter the data around a certain 
date, plus a collection of sinusoidal
functions, while for wavelet analysis the wavelet functions (Eq. \ref{morlet6-eq}) play the role both the filter
and the (complex) periodic function. 

\begin{figure*}
\centering
\includegraphics[width=0.8\linewidth]{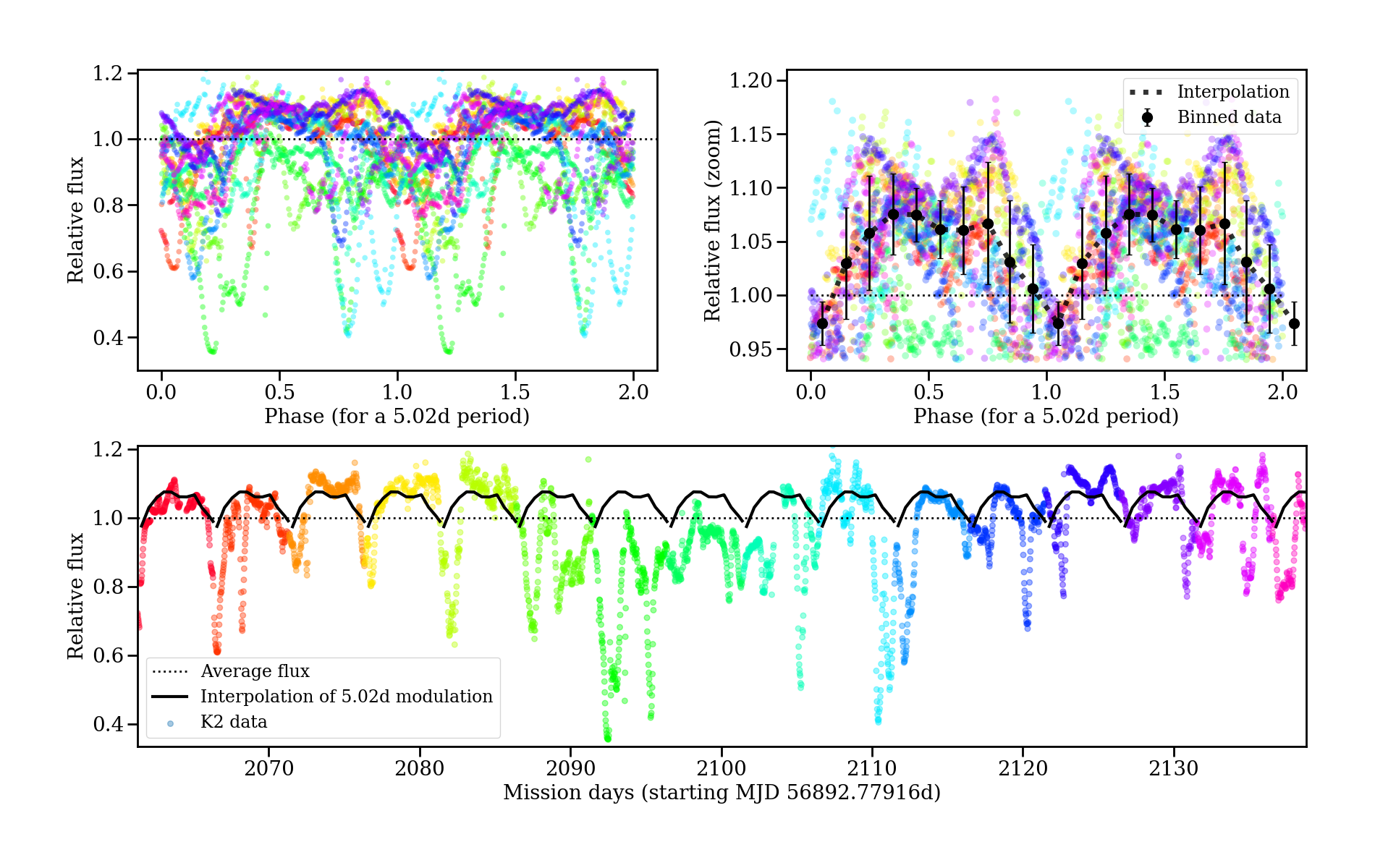}
\caption{K2 data wrapped according to a period of 5.015d. The data are color-coded according
to date to better display the points that were taken nearby in time. The upper left panel shows the entire K2 data wrapped, showing the phase twice to aid the eye. The upper right panel offers a zoom in the "out-of-eclipse" part of the K2 data (colored points, selected via $\sigma$-clipping), together with the average and standard deviation in 10 intervals in phase space (black points; the errorbars correspond to the standard deviation within each bin) and an interpolation curve to trace the shape of the variations. The bottom panel contains the full K2 data plotted against mission day together with the interpolated modulation curve of the top-right panel, to show how the phase of the eclipses drifts at certain times during the observation campaign, although the 5d periodicity remains visible throughout the full dataset.\label{K2wrapped-fig}}
\end{figure*}

The wavelet analysis
was performed using the Python PyCWT spectral analysis module\footnote{https://pycwt.readthedocs.io/en/latest/}
\citep[based on][]{torrence98,grinsted04,liu07}.
The procedure requires the data to be uniformly sampled, which is the case for K2
observations. The few small gaps and inhomogeneities in the data were filled by interpolation.
While this may have an effect on the shortest timescales sampled, it does not affect the 
final significant periods, which are in the range of days.
The significance was estimated considering a model for red noise 
estimated through a Lag-1 autocorrelation of the original data \citep{torrence98, liu07}. 
The results of the wavelet analysis are plotted in Figure \ref{wavelet-fig}. The wavelet
analysis recovers the results of the SGLSP and shows the same trends of quasi-periodicity,
with significant signatures in the range of 2.4-2.6d, 4.9-5.5d, and, to a lesser extent, 9d. 
As for the SGLSP analysis, not all the periods are recovered on all epochs and there is a 
drift in phase and periodicity that suggests changes in the structure that causes the eclipses.

\begin{figure*}
\centering
\begin{tabular}{c}
\includegraphics[width=0.8\linewidth]{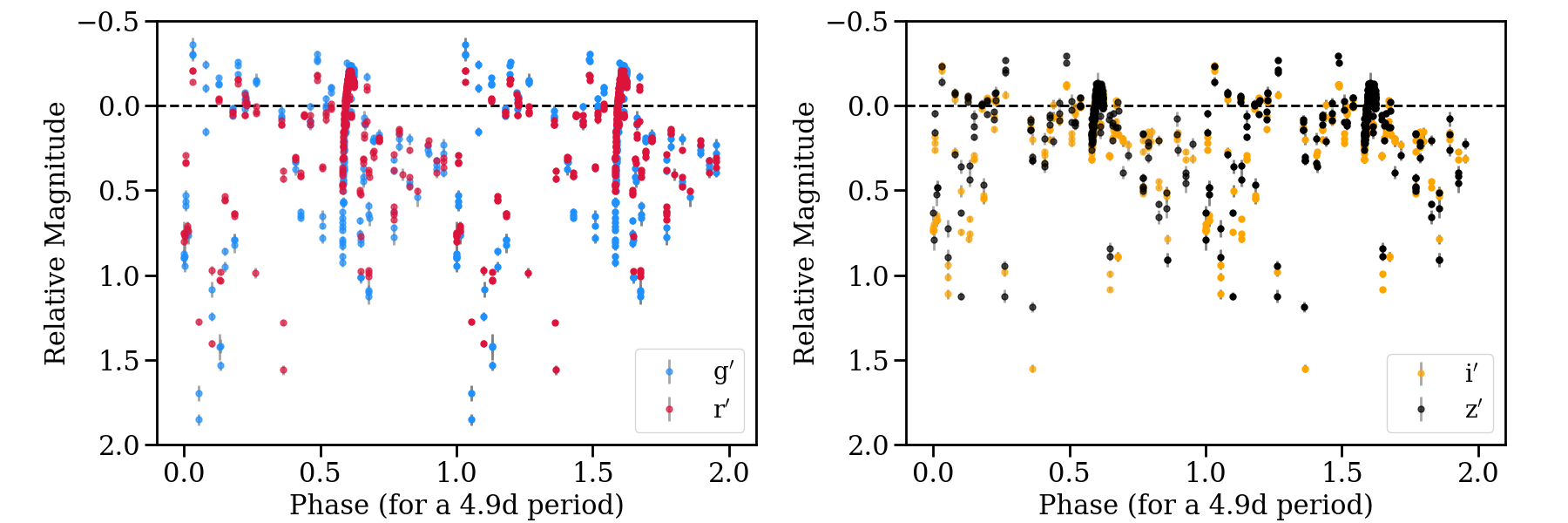} \\
\includegraphics[width=0.8\linewidth]{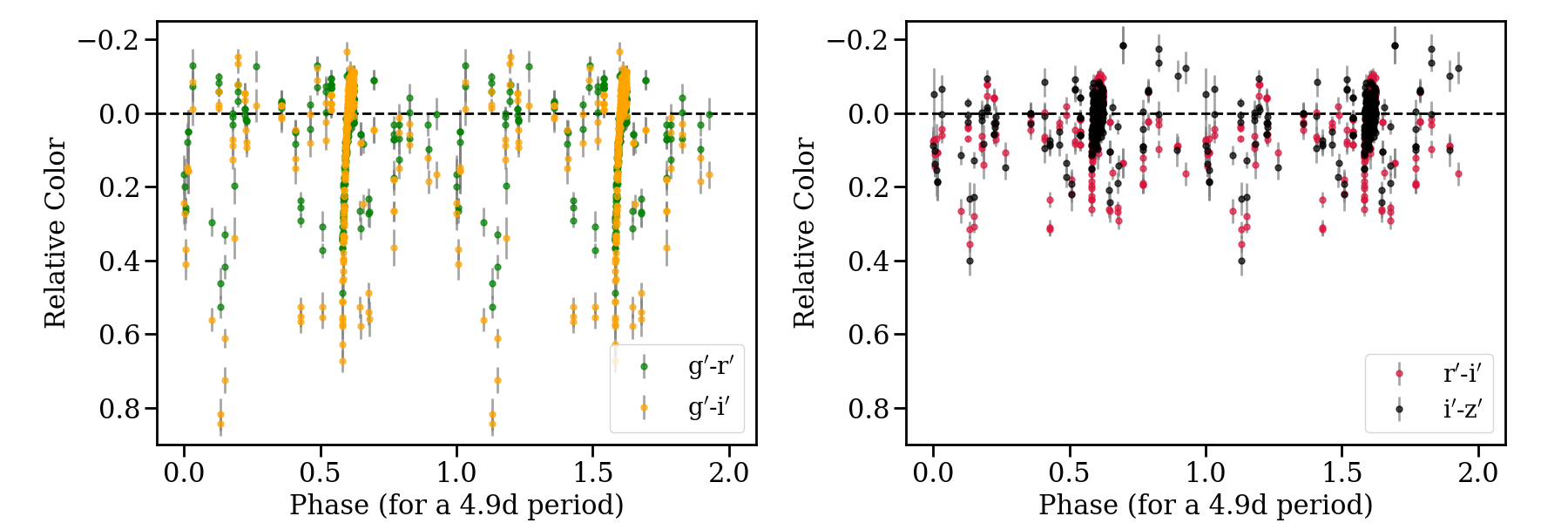} \\
\end{tabular}
\caption{
Top panels: REM data, wrapped for a 4.9d period to reveal the presence of periodic signatures.
Bottom panels: REM colors, wrapped for a 4.9d period. All the magnitudes are given relative to the median
magnitudes in each filter. \label{REMperiods-fig}}
\end{figure*}

To help visualizing the periodicity, we phase-wrapped the
data for the various potential periods. This exercise reveals a clear
modulation for a period of 5.02$\pm$0.12~d (corresponding to the main peak of the GLSP, see 
Figure \ref{K2wrapped-fig}) and, to a lesser
extent, for a period of half this value. Any other period fails the visual check or appears to be spurious
(e.g. being an integer multiple of the sampling rate).  The overall shape of the curve suggests a 5d
quasi-periodicity for the eclipses, and at the same time
reveals a clear but rapidly changing modulation in the flux observed out of eclipse.
The tip of the phased lightcurve has a "M" shape that could be consistent with rotational modulation
in a star with two non-identical cold spots (see Figure \ref{K2wrapped-fig} top right). 
Nevertheless, we also observe that the modulation suffers significant
changes from one period to the next, which is unfeasible for  typical long-lived, cold stellar spots. 
The eclipses are clearly associated to this 5d period, although they also change in depth from period to period and
show a drift in phase  during the observed epochs (Figure \ref{K2wrapped-fig} bottom). The 
epochs of increased eclipse activity correspond to the times when the 2.5d
period is stronger. Having ruled out
relatively stable structures (e.g. spots) for the variability, the observed lightcurve requires
something that changes on the timescale of the rotational period, such as significant variations in
the obscuring material in the innermost disk. The various possibilities will be discussed in
the following sections.

The REM data do not offer the same kind of time coverage and photometric stability, and thus 
the periodicity that can
be inferred from them has low-significance. Moreover, the correlated errors in the REM magnitude are 
very complex as the
"redness" of the noise strongly depends on the observed cadence and the magnitude (given the typical cadence, it is unlikely to find many consecutive points on eclipse so that low magnitudes tend to be followed by quite uncorrelated ones, while "out-of-eclipse" points are often followed by a measurement with very similar value). This makes it very hard to
assess the significance of the GLSP and makes a wavelet analysis impossible. Nevertheless, the 
same rough behavior is observed, and wrapping the lightcurve reveals a dominant periodicity around 4.9d that appears 
correlated with the extinction events (see Figure \ref{REMperiods-fig}). There is no high-cadence periodic signature.

\subsection{REM data, extinction,  and the inner disk structure \label{extinction-sect}}

The multi-band REM data allows us to study the color variation during the eclipses
for the first time. The JHK data could provide a good insight about the properties of
the obscuring matter, but the only dates for which we have 
complete data do not reveal substantial variability, and the only filter
for which we cover a significant number of points and shows variability is H. 
The optical data taken around the same dates as JK reveal that the object was essentially out of eclipse. 
The observed JHK colors are consistent
with the colors of a pre-main-sequence star without significant near-IR excess \citep{KH95}.

\begin{figure*}
\centering
\includegraphics[width=0.95\linewidth]{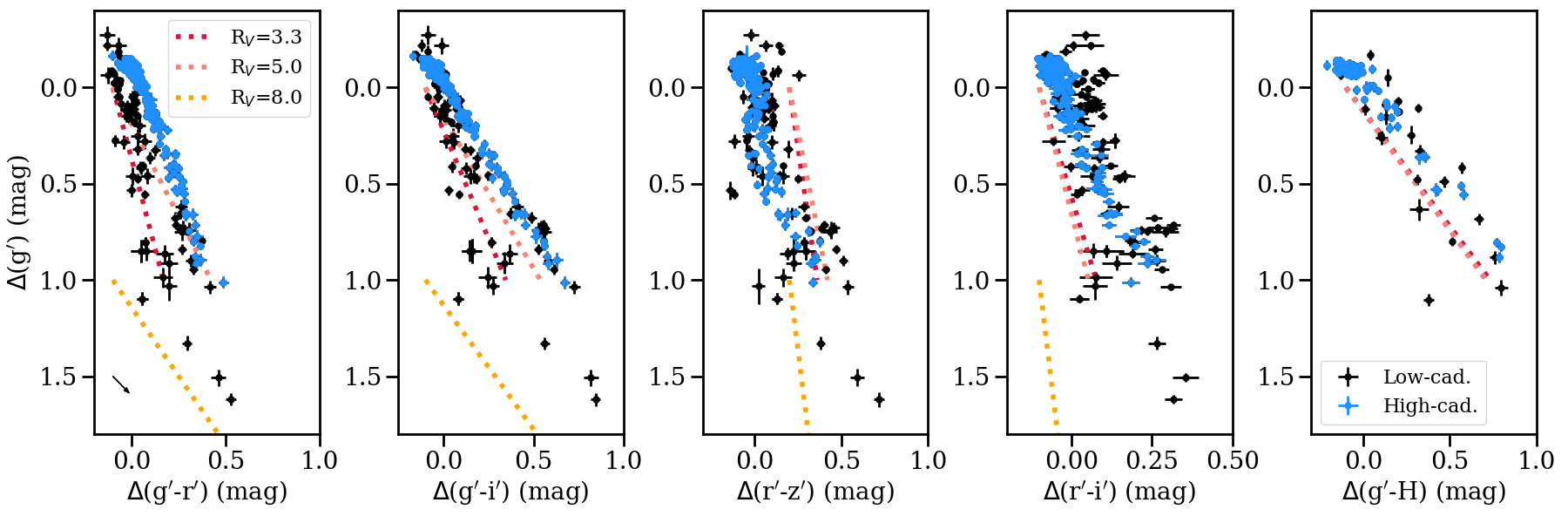}
\caption{Relative color-magnitude variations in the optical and H-band REM data. The dotted lines indicate the extinction vector for several extinction laws \citep[R$_V$=3.3,5.0 and 8.0, derived following][]{schlegel98}. High-cadence data are plotted in blue, while the rest of data are shown in black. All magnitudes are relative to the median value in each filters.  The black arrow in the left of the first panel
shows the correlation direction for cases where g' is also used to calculate the color, showing 3$\times$ the average uncertainty.
 \label{colormag-fig}}
\end{figure*}

We thus concentrate on examining the optical and H-band data. Figure \ref{colormag-fig} shows the color 
variation as a function of the magnitude variation for several combinations of bands. 
Considering the standard interstellar extinction laws \citep[R$_V$=3.1-5.5, see for instance][]{schlegel98,stoughton02, cardelli89}, the slopes of the color variations up to $\Delta$g'$\sim$1 mag 
are quite consistent with the standard extinction vectors.  There are some places where the color variation becomes suddenly smaller, for instance around $\Delta$g'$\sim$1 mag and (especially in the g' vs r'$-$z' diagram), around $\Delta$g'$\sim$0.2 mag. These could be due to
scattering shifting the colors towards bluer regions as the eclipse
progresses \citep[as has been observed in UXors, e.g.][]{grinin88}.
There are further changes in the slope as the eclipse progresses, with the deeper eclipses being
better fitted with an extinction law with higher R$_V$ \citep[which is in general attributed to high extinction clouds
with more processed material;][]{cardelli89} and the very deep eclipses having an offset
with respect to the standard trend and showing a flatter reddening curve.

The color variation towards flatter reddening curves is only observed for eclipses with
 $\Delta$g'$\sim$0.9-1.1 mag. In those cases,
the reddening continues into the deepest eclipses (for which $\Delta$g'$\geq$0.9)
with a slope that cannot be explained with the standard R$_V$=3.1-5.5. The slope variation 
 is likely related to strongly processed
dust grains, although the change is different in different filters, maybe indicating an
effect of grain size \citep{eiroa02}.  Higher density or differences in dust properties and/or 
scattering within the clumpy material of the disk could also contribute.

There are also some shifts parallel to the extinction vectors visible at about all magnitudes.  Small variations in the stellar luminosity (e.g. due to accretion) 
and/or scattering on longer timescales may also contribute
to shift some of the data taken on different dates parallel to the standard extinction vectors
as seen in Figure \ref{colormag-fig}, even though small stellar luminosity and accretion changes are not
expected to produce a significant change in the observed colors. Note that a large fraction of the  shallower eclipse points belong to the high-cadence observations and thus were taken
very close in time,  so that their variation is smooth.

Leaving aside the UXor behavior towards bluer color, the changes in the color slope 
consistent with reddening could be caused by  extinction by 
more opaque material in the deepest parts of the eclipse and  
variations in the sizes of the grains depending on the height in the innermost disk \citep[e.g.][]{mcginnis15}, 
due for instance to differential settling \citep[e.g.][]{dalessio06,laibe14}. 
The dust content around a star is constrained by the 
dust sublimation temperature \citep[$\sim$1500 K, with some variations in the 1000-2000 K range depending on disk structure, density, and composition;][]{isella05,kama09},  and typical observations can be
well-fitted with inner rims at T=800-1200 K \citep{mcclure13}. 
Significant grain processing happens already at these (and much lower) temperatures \citep{tielens05}, so that 
the grains are likely different from plain ISM silicates. Grain growth is also generalized in disks, 
and larger grains tend to 
produce grey extinction with a lower color dependency at optical and near-IR bands \citep{miyake93,eiroa02},
and large grains are often involved in the best-fitting models for inner disk walls \citep{mcclure13}.
In addition, there are other effects such as
a dusty wind \citep[e.g. as observed in RW Aur;][]{bozhinova16b} that could also cause obscurations. The H$\alpha$
profile of RX~J1604.3-2130A shows some blueshifted absorption (Manara et al. in prep., 
see also Section \ref{variability-sect}), but
since the accretion rate is low,
accretion-related winds are expected to be weak and carry significantly less mass than the
accretion flows.

\begin{figure}
\centering
\includegraphics[width=0.95\linewidth]{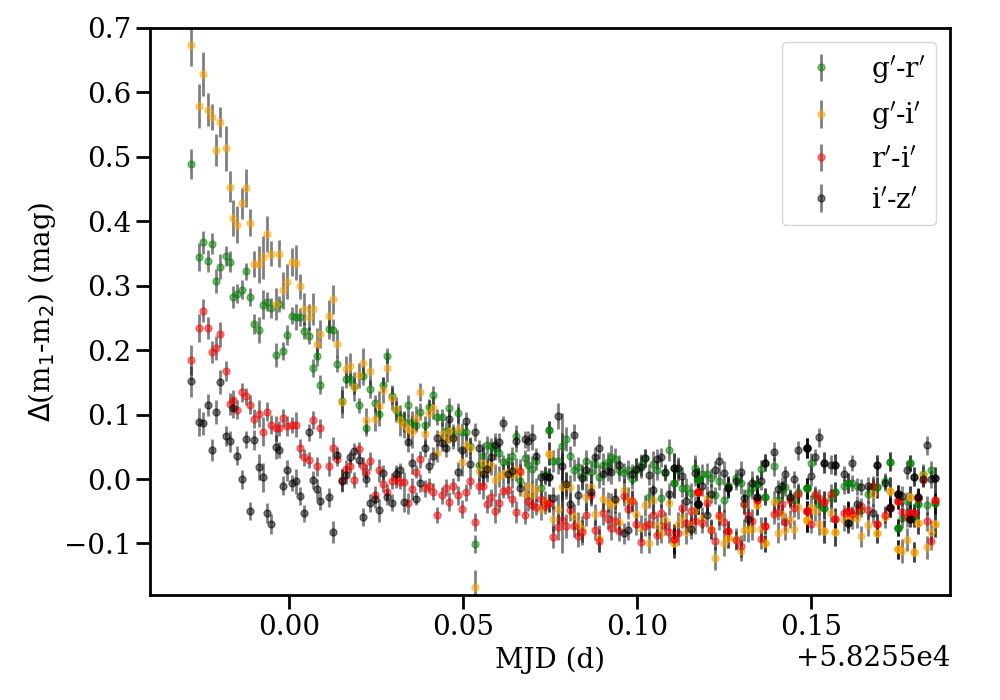}
\caption{Color variations in the dip observed with the high-cadence data. Note that the eclipse is not deep enough to show significant variations in the extinction law, although some of the UXor behavior is seen,
especially in i'-z'.\label{coloreclipse-fig}}
\end{figure}

The high-cadence data offers us a chance to explore what happens during a single eclipse on a timescale
where the variations (accretion, luminosity) are likely negligible (Figure \ref{coloreclipse-fig}).
The data show a smooth transition 
from eclipse to maximum,  although the high-cadence eclipse does not go as deep as those where a
significant color offset is observed. The  high-cadence data are fully consistent with a mild UXor behavior around $\Delta$g'$\sim$0.2 mag, plus increasing dust
extinction with standard extinction laws for thin ISM dust (R$_V$=3.3).
For the rest of the eclipses, including those happening at half of the period, the
coverage is very scarce, but wrapping the data does not reveal a significant difference between the colors of
different eclipses, nor between the colors of the full-period vs half-period eclipses (see Figure \ref{REMperiods-fig} bottom), other than the differences observed between shallow and deep eclipses.

\begin{figure}
\centering
\includegraphics[width=0.95\linewidth]{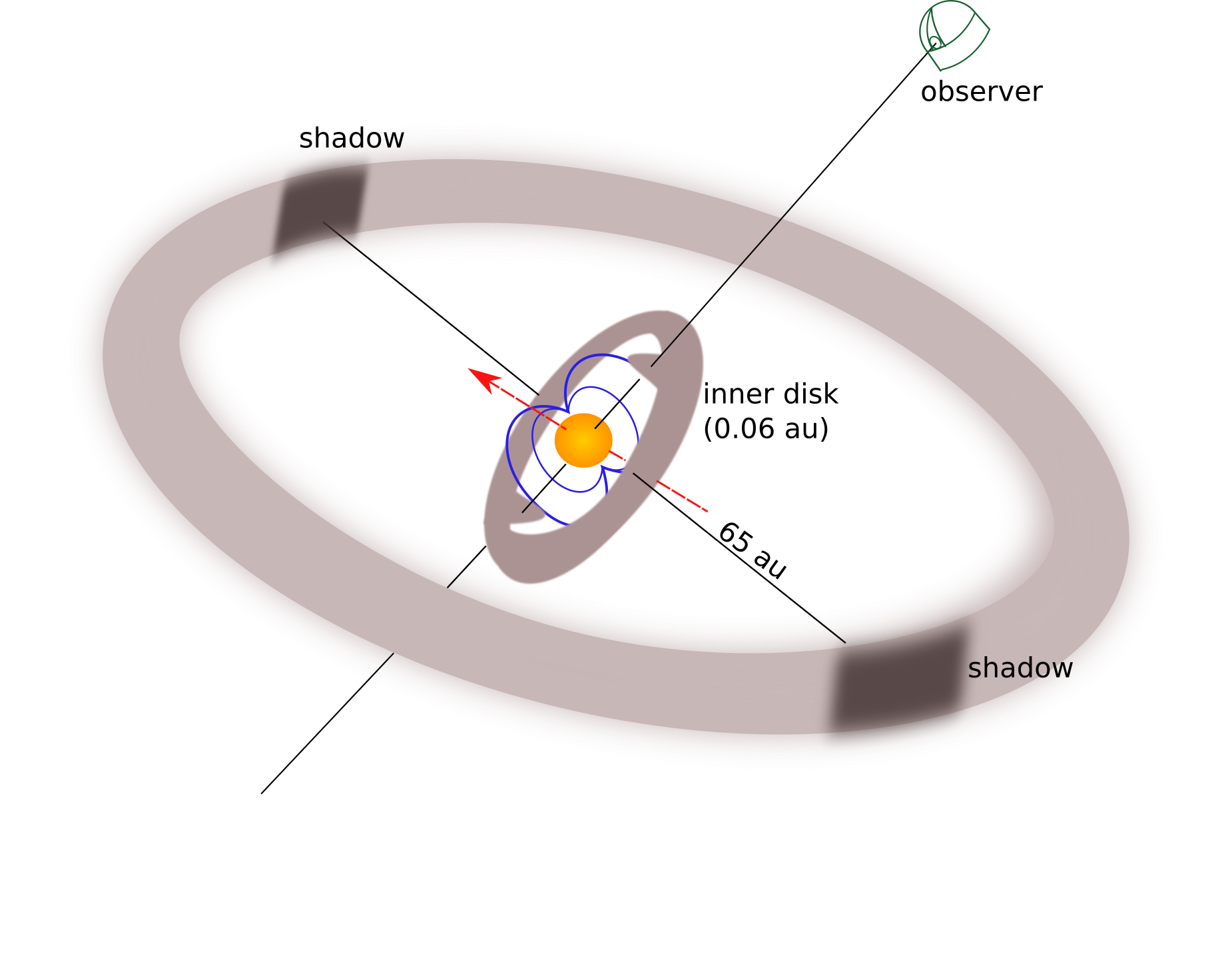}
\caption{Not-to-scale sketch showing the inferred structure of the system. The star and its magnetosphere are
displayed in the center, the red arrow indicates the rotation axis of the star, magnetosphere, 
and corotating inner disk. The
inner disk is deformed in the regions where the magnetosphere is joining in, which causes warps that cross our line-of-sight when the disk rotates. The outer and inner disk are highly inclined with respect to each
other, so that there are always shadows along the line where the plane of the inner disk and the plane of the outer
disk cross. Note that the inner disk is expected to be more wobbly and irregular than shown here.
 \label{cartoon-fig}}
\end{figure}

We can then use the observed extinction to estimate the amount of material needed to produce the 
eclipses. The typical depth of the extinction event in g' is 1.2 mag. Assuming standard
interstellar extinction \citep{schlegel98}, this is equivalent to A$_V$=1 mag, or
N$_H$=1.8e21 cm$^{-2}$ \citep{prehdel95}. For an average particle weight of 1.36 m$_H$  \citep[for solar metallicity neutral gas, e.g. see][]{mihalas78}, we obtain 
0.004 g cm$^{-2}$.  Note that because the metal vs hydrogen content in the inner disk may be different
\citep[either higher or lower, e.g.][]{panic09,riviere13}, 
there is some uncertainty in this parameter. The stellar parameters suggest a dust destruction radius at 0.06-0.15 au.
If this mass were distributed in a ring of radius 0.06 au (0.15 au) and height comparable
to the stellar radius, the total mass associated to the structure would be about 2e-3 (5e-3) M$_{Ceres}$,
including gas and dust. Nevertheless, 
we can derive a better estimate of the thickness of the structure from the observed shadows in the
outer disk, that span about 20 deg in average (see sketch in Figure \ref{cartoon-fig}). For an obscuring structure at 0.06 au (0.15 au) distance,
this would mean a size about 0.02 au (0.05 au) or 3R$_*$ (8R$_*$) as a function of the stellar radius.
The total as plus dust mass would be in the range 1-6\% of the mass of Ceres.
This means that the innermost disk ring does not need to be very massive in order to reproduce the
observed behavior. An extinction law for dark nebulae would result in a slightly larger mass, although for 
reasonable values the main uncertainty in our estimate remains the uncertainty in the size
of the innermost disk and the gas to dust ratio.

\citet{pinilla18} revealed flux variations in the scattered light J-band 
observations up to 0.4-0.6 with respect to the non-obscured flux. Standard
extinction laws suggest A$_g \sim$2.6-4.3 mag,  about a factor of 1.4-2.3
higher than in the deepest optical eclipses, which would result in a similarly larger mass. 
There are several possibilities to explain this difference. Differences in alignment
between the star-inner disk and the inner disk-outer disk could result in different column densities viewed on each line of sight (see Figure \ref{cartoon-fig}). In addition,
the fact that the shadow observations and the optical photometry are not simultaneous and the dust content
is known to be variable, plus the possibility that
the dust does not follow a standard extinction law, especially in the deepest eclipses, may
also play a role. Other
possibility would be if the inner disk does not cover the full stellar disk
along our line-of-sight, or
if the disk does not generally covers the surface of the star but a localized warp or blob on it does.
In fact, the observed differences in extinction (observed from the lightcurve 
A$_g \sim$0.4--1.8 mag, expected from shadows A$_g \sim$2.6-4.3 mag) can be explained with a 
variable stellar disk coverage  by the eclipsing material. Considering the density estimated from the extinction
a variation in coverage ranging from 0\% out of eclipse, to
between 30\% for shallow eclipses, down to 100\% at the deepest ones could explain the observed lightcurve.
Note that the optical variability depends on various poorly-constrained parameters, such as the
relation between infrared and optical extinction (i.e., dust properties), whether the
disk is occulting the star towards the equator or towards the poles (due to limb darkening), and whether
the star has additional causes of variability (e.g. hot and cold spots). 

The observed small changes and displacements of the
shadows are consistent with a "wavy" or warped disk \citep{pinilla18}, 
which is also consistent with other observations \citep{grinin08,mcginnis15}, models 
of self-shadowed UXor disks
\citep{dullemond03},  and the non-axisymmetric or clumpy structures 
observed in gas and dust in the inner disks of some young stars \citep[e.g.][]{sicilia12,siwak14,scholz19}.  A warp 
at the point where the disk and the star are connected at the basis of the accretion column \citep[see e.g. the models of][]{alencar12,alencar18,mcginnis15,bodman17} could also explain the observations,  as shown in the
cartoon in Figure \ref{cartoon-fig}.  The variations of the position of the
shadows \citep[$\sim$20 deg in average;][]{pinilla18} together with the estimated radius of the disk 
suggest that the innermost disk is as
wavy as it is thick. A wavy/warped disk that deviates from a flat structure in vertical scale by
about a third of the disk radius (to account for the angle variation of the shadows) would also explain
why the star is not always eclipsed, while the shadows on the outer disk are nearly ubiquitous despite variability.

A 5d orbital period corresponds roughly to the distance where the disk would have a temperature around 
1540 K ($\sim$0.06 au or 9.4 R$_*$). The
precise location and temperature depend strongly on the 
dust properties, on the structure of the inner disk rim, and on the
stellar parameters (especially, the stellar mass derived from evolutionary tracks). 
But in any case, the corotation radius is compatible with
the dust sublimation radius.
With this in mind, the quasi-periodicity of the eclipses discussed before suggests that the 
structure causing them is rapidly changing on timescales comparable to the Keplerian period of
the material in the innermost disk (days), so that even consecutive eclipses have different
depths and lightcurve profiles. Disk precession may alter the inclination of the inner disk
in time, but their timescales are typically much longer.  
For the general relativistic precession, the timescale is proportional to $(c/v_K)^2$ 
times the orbital period, where
c is the speed of light and $v_K$ is the Keplerian orbital velocity. Kozai or 
secular resonances may also
cause precession, but they are usually weaker than the relativistic effect \citep[][]{kozai62,ford00} and 
require very massive and close-in companions that would have been spectroscopically detected .  

The quasiperiodicity on short timescales and color variations suggest that
the disk is highly asymmetric, with warps or clumps that are denser than the rest. It is also likely being
externally modified/fed due to viscous matter transport, which can explain the sudden periods of intense dimming and 
concatenated eclipses, such as the one observed
between days 2085 and 2105 in the K2 data (see Figure \ref{K2lightcurve-fig}). 
Taking into account the accretion rate  $\sim$3e-11 M$_\odot$/yr
and assuming that the rate of transport of matter in the inner disk is similar, the mass of the
innermost disk required by the eclipses is comparable to what accretion transport could provide in 
between two weeks to two months time.
This means that the inner disk is filling up (and draining) on relatively short timescales, so that significant
changes could be observed on 5d timescales.

To summarize this section, the color variability confirms that the eclipses are consistent with
extinction by dust with properties ranging from ISM dust to more processed grains, located in an irregular, warped 
disk at the corotation
radius. The total mass depends on the dust properties and on the size of the disk, but $\sim$1\% M$_{Ceres}$
of total mass (including gas and dust)
is enough to explain the observations.
Considering the accretion rate, it is not unexpected that
the dust distribution in the innermost disk changes on short timescales, since a significant fraction of the
obscuring matter in the innermost disk will be  fed to the star (or
drained from the innermost disk) on each rotational period, explaining the rapid variability.

\subsection{Rotation, accretion, and variability in the inner disk\label{variability-sect}}

\begin{figure}
\centering
\begin{tabular}{c}
\includegraphics[width=0.8\linewidth]{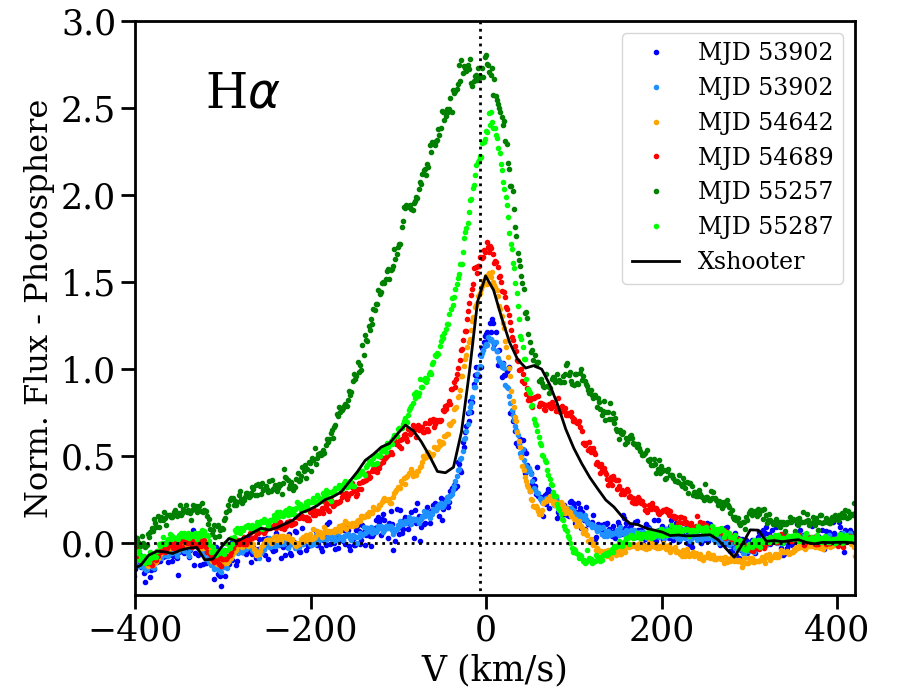} \\
\includegraphics[width=0.8\linewidth]{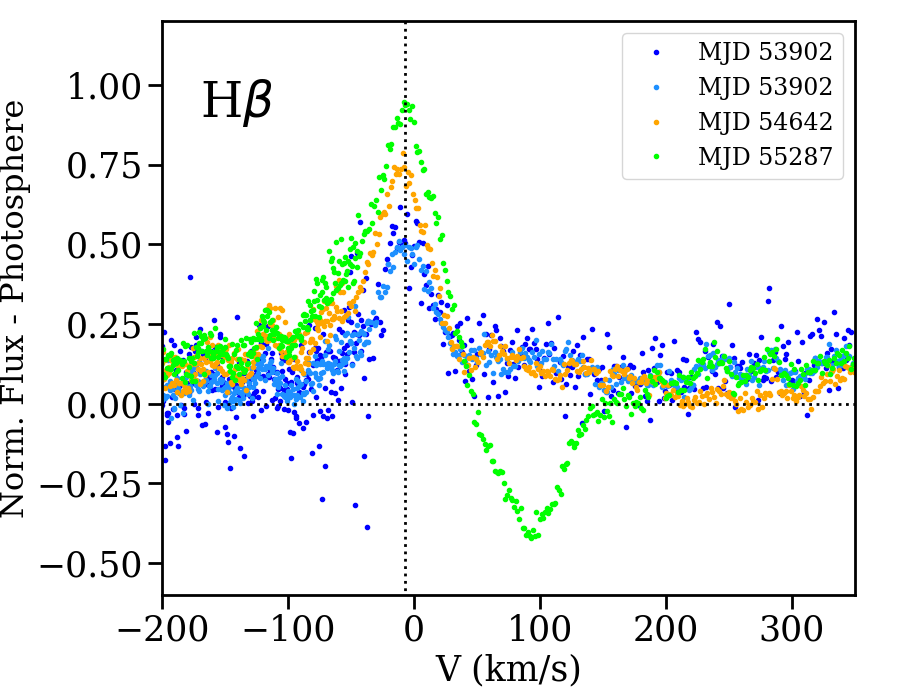} \\
\end{tabular}
\caption{Photosphere-subtracted H$\alpha$ (top) and H$\beta$ (bottom) spectra observed with HIRES/Keck
(for various epochs) and X-Shooter (corresponding to the time when the accretion rate and the stellar properties were estimated; Manara et al. in prep). 
The zero level for flux and radial velocity are marked by dotted lines. A rotationally-broadened template photosphere has been subtracted to show the absorption and emission features.  \label{halpha-fig}}
\end{figure}

Rotation rates suggested a very high inclination for the disk system around RX~J1604.3-2130A \citep{davies19},
despite its outer disk being nearly face-on \citep[6 deg;][]{zhang14,dong17}.
If the 5d modulation observed was caused by rotation, we can use the observed rotational
velocity (16.2$\pm$0.6 km/s) to infer 
the inclination of the system.  The maximum period that this velocity range allows is 4.2-4.5d. 
A 5d period would require
a stellar radius $\sim$10-15\% larger than estimated, or larger uncertainties in the v$sini$ (e.g. due to 
differences in limb darkening). 
In any case, the star cannot be aligned with the outer disk because it would rotate at breakup velocity,
but should be closer to equator-on and thus rather aligned with the inner disk.
If the 2.4-2.5 d period were a rotation
period, the inclination angle would be about 38 degrees (still far from aligned with the outer disk). Nevertheless, such a low angle would be inconsistent
with the lower limit of 61 degrees of \citet{davies19}, besides the fact that the shape of the lightcurve 
agrees rather with a 5d rotational period in a star with two asymmetric eclipsing structures, rather than with a 2.5d period.

\begin{figure*}
\centering
\includegraphics[width=0.8\linewidth]{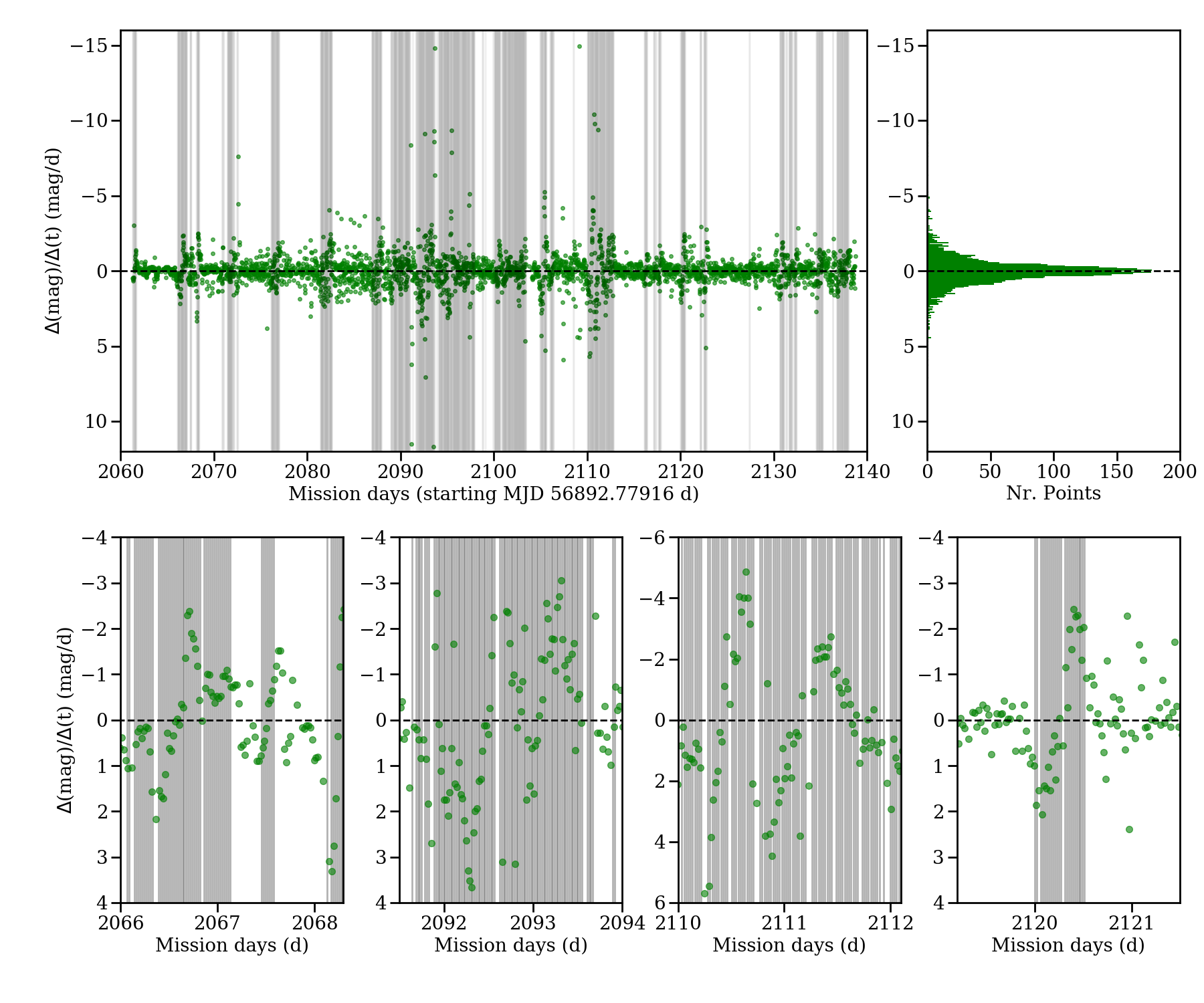}
\caption{Time variation for the K2 observations. The upper left panel shows the change in magnitude per day as a function of the observing date Grey lines mark the cases in which the K2 relative flux drops below 0.94 (the limiting value for eclipses considered, see text). The upper right panel is a histogram of the same plot. The lower panels are a zoom-in of the magnitude change vs time for several of the eclipses. \label{magvelo-fig}}
\end{figure*}

Constructing a model for the out-of-eclipse lightcurve is complicated, because the 
sampling of the REM photometry is not high enough to show the detailed color variation during a single rotational period and
there are significant changes on a few-days basis (see Figure \ref{REMperiods-fig}). 
A cold spot model is not sufficient since spot-related
variability is usually of the order of 0.1 mag \citep{grankin07} and can by no means explain the observed eclipses
nor shadows, but it could explain the M-shaped part of the K2 data. 
The flux modulation at phase $\sim$0.5 (see Figure \ref{K2wrapped-fig}) can be obtained with spots 200-500 K cooler than
the stellar photosphere and spot coverage between 0.03 and 0.36 \citep[as in e.g.][]{bozhinova16}, all of them reasonable parameters, but the rapid
variability from period to period is hard to explain.
For instance, a smaller-scale eclipse could also produce the same effect, and since the amount of matter responsible 
for this would be very small
compared to what is observed to flow around the star from its accretion rate, rapid variations are more plausible.
In addition, there is more evidence of the disk-star connection being dynamic and rapidly variable than in the case of 
stellar spots \citep[e.g.][]{fonseca14}, and the color variations (Figures \ref{REMperiods-fig}, \ref{colormag-fig}, and \ref{coloreclipse-fig}) suggest that occultations are the dominant process.

All the datasets (CSS, K2, and REM) show brief periods of increased eclipsing activity on 
timescales shorter than 5d (usually, about 2.5d).
These could be times at which an increased accretion rate throughout the disk triggers accretion onto the other side of the star and feeds a secondary warp, causing additional
extinction events. Modeling this situation would require, at a minimum, simultaneous high-cadence 
multi-color photometry and spectroscopy 
and it is thus beyond the scope of this paper.

The archival HIRES data are consistent with
the low accretion rate estimates from X-Shooter spectra (Manara et al. in prep), not 
showing the emission
lines characteristic of strong accretors \citep[][for instance, no He I emission and Ca II IR 
narrow emission lines in the center of strong absorption components]{hamann92}. 
There is also no evidence of spectroscopic binarity (see Appendix \ref{kinematics-app}).
H$\alpha$ and H$\beta$ have
strongly variable intensity and line profiles, with timescales of months to years.
The features are better viewed after photospheric subtraction, using a 
photospheric template derived from the standards HD 114386
(which has been observed in the H$\beta$ region) and HD 151541 (for which the available data covers the H$\alpha$ region; 
see Figure \ref{halpha-fig}).
After photospheric subtraction, H$\beta$ emission becomes evident, together with 
evidence of both redshifted and (in H$\alpha$) blueshifted absorption components. 
Photospheric subtraction also reveal further complex absorption profiles in other lines, such as Na I D. 
Since there is no detailed day-to-day high-resolution spectroscopic followup, it is hard to assess whether the
variations in the line profile are due to variations in the accretion rate or changes in the orientation
of the accretion flows with respect to the observer. Nevertheless, the changes 
in equivalent width and 10\% velocities suggest that
the accretion rate is variable, with up to 2 orders of magnitude variability between the maximum and the minimum
width \citep[using the 10\% H$\alpha$ width;][]{natta04}, with the 3e-11 M$_\odot$/yr value from Manara et al. in prep. being an intermediate
rate. Accretion is very weak (consistent with no 
accretion) in the spectra from MJD 53902. 

For both the H$\alpha$ and H$\beta$ lines, the redshifted absorption features are
dominant, and on two of the dates (MJD 54642 and MJD 55287) show a YY Ori 
or inverse PCygni profile  (IPC) with
an absorption component that goes below the continuum. This suggests that the accretion column
must have been very close to viewed along the line-of-sight  on these dates. Accreting along the line-of-sight could
also mean that the gaseous material could
be also obscuring the star \citep[as has been suggested in other objects, e.g. TW Hya;][]{siwak14}. Obscuration
by dust could happen in a warp induced in the place where the accretion column is attached to the disk \citep{mcginnis15,alencar18} although unfortunately
none of the spectra has simultaneous photometry. The spectrum taken on
MJD 54642, which shows a mild IPC profile in H$\alpha$, is the one closest to a photometric datapoint from CSS 
corresponding to a deep eclipse. However, the HIRES data were taken 24h after the photometry, and
since typical eclipses last less than 24h, it is not possible to assume that the star would have been in eclipse. 
Moreover, phase shifts are expected between the very-close-in gas emitting H$\alpha$ and 
eclipses associated to dust at the corotation
radius. On MJD 55287, the IPC is clearly visible in
H$\alpha$ and very strong in H$\beta$, a typical signature of highly inclined system  \citep{alencar12, alencar18,donati19}.
This, together with the observed eclipses, is a sign that RX~J1604.3-2130A may be very similar to other
highly-inclined systems such as Lk Ca15, AA Tau, and V354 Mon \citep{bouvier07,alencar18,donati19,fonseca14}.

Since the analysis of the extinction suggests an inner disk that is routinely drained on a short timescale due
to accretion, we explore whether the rate at which the magnitude varies during the eclipse is consistent with
material transported due to accretion. A change in magnitude vs time is 
equivalent to a change in column density over
time, using the conversion between extinction and column density as in Section \ref{extinction-sect}.
Assuming that this extinction event covers the stellar surface uniformly, the 
obscuring mass involved per time can be obtained by
multiplying by the area of the stellar disk. This value can be then 
transformed into a approximate "projected" accretion rate
that can be compared to the accretion rate measured by spectroscopy. Because the REM
data has only very scarce coverage of each dip, we need to do the exercise with the K2 data, although one of the main
limitations is the lack of color information. 

For K2, we calculate the change in magnitude between each two points 
$i$ and $i+1$ using the 
flux ratio between these two points to estimate the variation in 
magnitude as $-2.5 log_{10} (f_{i+1}/f_i)$. Dividing by the
time interval we can obtain the change in magnitude vs time (Figure \ref{magvelo-fig}) and transform it into an 
approximate accretion rate as explained above. Using the conversion between K2 extinction and A$_V$ 
\citep[A$_V$=0.4 A$_{K2}$\footnote{Note the value is approximate as 
it depends strongly on the color of the source.};][]{rodrigues14}  we obtain a typical change of 1 mag/day. With 
the relation between A$_V$
and column density \citep{prehdel95}, we derive an approximate accretion rate of
6e-12 M$_\odot$/yr. The large majority (99.1\%) of the observed points fall between the (-5,+5) mag/day interval, and
these would correspond to accretion rates up to 3e-11M$_\odot$/yr, fully consistent with the accretion rate estimates 
from X-Shooter observations (Manara et al. in prep). Note that these "changes" do not need to 
be accretion rate variations, since we are only taking into account matter along the line-of-sight,   and that the
estimates are also subject to the same uncertainties than the inner disk mass estimates, namely dust properties 
and gas to dust fraction.
It is also important to keep in mind that the obscuring matter contains dust and must then be located at least at
the dust destruction radius near the star-disk connection, while what is observed in
H$\alpha$ corresponds to hot gas likely closer to the stellar photosphere. Because of this, 
changes in the dust are not expected to be observed as immediate changes in H$\alpha$, and a phase
delay is very likely to occur, as it is also observed between  accretion-related emission lines with different energetics \citep{sicilia15}, or between veiling and line emission, or optical and X-ray accretion signatures \citep[][]{dupree12}.

\subsection{The  long-term variability of the inner disk \label{diskIR-sect}}

\begin{figure*}
\centering
\includegraphics[width=0.7\linewidth]{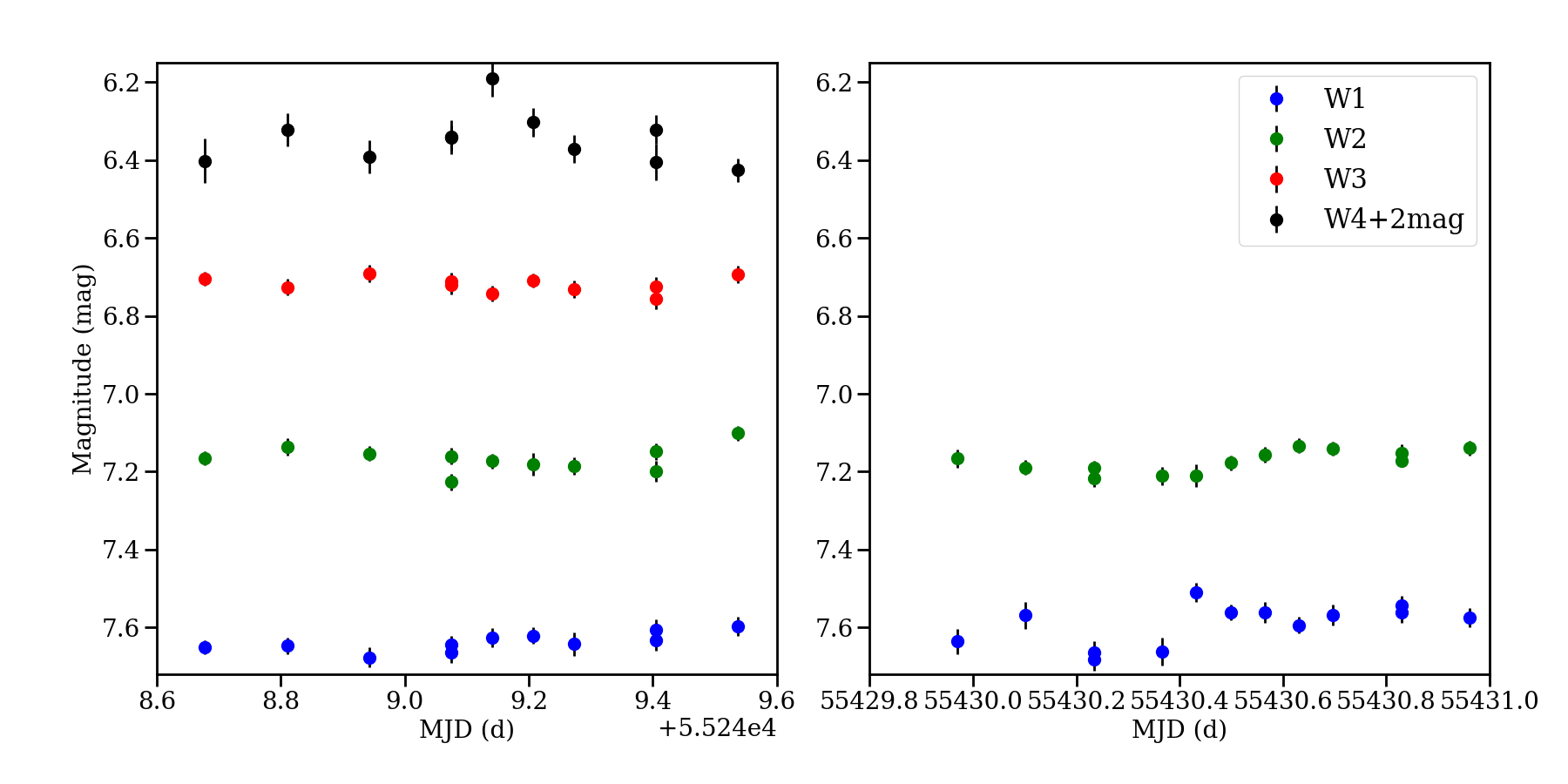}
\caption{WISE lightcurve during the two periods of observation. The W4 magnitude, significantly brighter than the rest, has been shifted by 2 mags for display. \label{WISE-lightcurve}}
\end{figure*}

The disk around RX~J1604.3-2130A has been classified as transitional \citep{carpenter06, mathews12,esplin18}.
\citet{luhman12} pointed out that the mid-IR observations with Spitzer/IRAC were
consistent with a bare photosphere, while WISE data revealed a clear IR excess. Considering the fluxes provided by \citet{carpenter06}
and the IRAC zeropoints\footnote{See IRAC Handbook,  https://irsa.ipac.caltech.edu/data/SPITZER/ docs/irac/iracinstrumenthandbook/17/}, we find that the source 
changed by about 1.47 mag at 4.5$\mu$m between the Spitzer and the WISE observations\footnote{Spitzer IRAC2 and IRAC4 magnitudes are 8.64 and 8.38 mag, respectively.},
which are roughly separated by 4 years time (from  MJD 53820 to MJD 55249). 
The AllWISE lightcurve, on the other hand, reveals only mild mid-IR variability (as expected if the cause is extinction)
during the half a year interval during which RX~J1604.3-2130A was observed by WISE (see Figure \ref{WISE-lightcurve}), with amplitudes 
of 0.17 mag (W1), 0.13 mag (W2), 0.07 mag (W3) and 0.24 mag (W4). It is interesting that while W1, W2 and W3  roughly 
follow the same pattern and vary in parallel,  W4 behaves differently, although the uncertainties are also larger.
At longer wavelengths, the extinction becomes negligible, so large changes in the flux are most likely related
to changes in the disk structure. 
Wavelengths around 22$\mu$m trace material at considerably larger
distances, and could be dominated by the emission of the outer disk, 
less variable on short timescales. Note that the dramatic ($>$1 mag) IR variability 
affects only the IRAC bands, since W4 and MIPS 24$\mu$m roughly agree
despite being observed at times where 
there is substantial difference for the 3-10$\mu$m fluxes.  Therefore, although the 
strong variability in the near-IR 
is reminiscent of the "seesaw" behavior observed in some disks \citep[e.g.][]{espaillat11,flaherty12}, we note that
the situation here is quite different, because the mid-IR fluxes are not variable. While seesaw behavior in
(usually, pre-transitional) disks can be
explained by changes in the vertical scale without much modification to the contents of the disk, here 
the observed near-IR photospheric colors need a very strong depletion of warm dust.

The CSS data reveal 
decreased eclipse activity soon after the Spitzer observations were obtained, 
which points towards the intriguing possibility that the inner disk may 
have been depleted of dust around that date. The 
CSS data closer in time to the Spitzer observations (56 datapoints in 
total within 10-20 days) do not show any eclipse, although there are 
no observations for nearly half a year before the Spitzer observations and the first data point was taken 
more than 40 days after Spitzer. The sparse sampling may have missed the eclipses, since the CSS 
observations were obtained at irregular intervals of a few days during 2 months. However, with the sampling frequency
and given that, altogether, 35\% of the CSS observations appear to have been taken during eclipses, 
it is significant that not a single eclipse would have
occurred near the time of the Spitzer observations, suggesting 
that the eclipse activity may have indeed been lower at the time.
In addition, the HIRES spectra taken during the CSS 
stable phase after the Spitzer observations show a significantly weaker H$\alpha$ feature, consistent with decreased 
accretion and maybe an additional signature of lack of material in the innermost disk.

\begin{figure}
\centering
\includegraphics[width=0.99\linewidth]{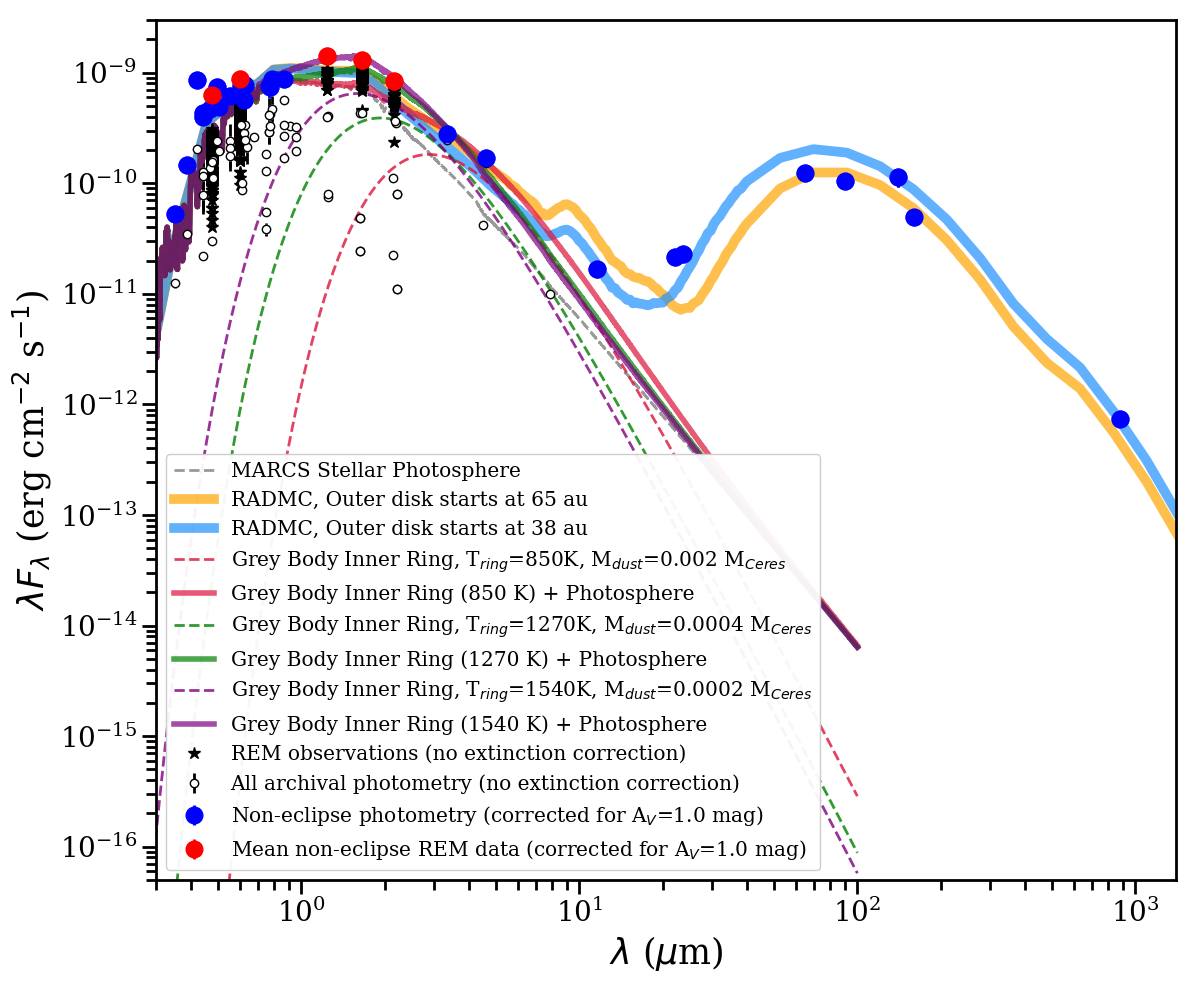}
\caption{SED with archival data for RX~J1604.3-2130A and simple disk models. The open dots show all 
the available data obtained from VizieR and not
corrected for extinction. The REM data are shown for the filters with negligible color terms.
The blue dots show the selected brightest magnitudes for 
each band, corrected by the nominal A$_V$=1.0 mag \citep{preibisch99}. Three very simple models are shown for
comparison for the purpose of estimating the mass that could be associated to inner disk emission (see text for details): Two constructed using RADMC radiative transfer code, and two assume a grey body model for material located in a
ring with temperature 850 K and  dust mass 0.003 M$_{Ceres}$, a temperature 1270 K,  dust mass 4e-4M$_{Ceres}$,
and a temperature of 1540 K with dust mass 2e-4M$_{Ceres}$,  respectively. \label{SED-fig}}
\end{figure}

The strong Spitzer vs WISE mid-IR variability is very similar to what 
has been observed for GW Ori, a triple stellar system with a circumtriple disk and
an unstable or "leaky" dust filter that leads to remarkable changes in the SED due to  variations 
in the innermost disk dust content 
on timescales of decades \citep{fang14}. 
Similarly to GW Ori, we can make an order of magnitude estimate of 
the amount of dust  needed in the inner disk to account for
the mid-IR variability.  This can be compared with the accretion rate and with the 
feeding rate needed to support the variability of the innermost disk structure deduced from the
extinction events, assuming matter flows through the disk in a stable way.
Nevertheless, studying the SED of RX~J1604.3-2130A is more complicated because the 
data are non-simultaneous and, unlike GW Ori, the star is highly variable in the optical.
To constrain the disk properties, we need to disentangle what is caused by extinction
from what may be caused from variations in the innermost disk structure. 
We construct the SED using VizieR multiwavelength data (see Appendix \ref{SED-app}) plus our REM 
photometry. We consider as stellar photosphere (out-of-eclipse data)
the brightest magnitudes observed in each optical filter. The out-of-eclipse data are corrected for A$_V$=1 mag
 \citep{preibisch99} using standard color relations \citep{cardelli89,stoughton02,bessell05}. 
The data fainter than the
out-of-eclipse values are not corrected for extinction and not used for the luminosity fit.
The SED reveals that the JHK
archival data were likely taken during one of the obscuration episodes (see Figure \ref{SED-fig}), so that
our REM observations are brighter than previous ones and more consistent with a K3 photosphere.
The CSS observations suggest that the star was in a bright state near the Spitzer 
observations, and the WISE observations do not
seem seriously affected by the strong optical variability at the time. Thus it 
is reasonable to assume that the main changes at wavelengths $>$4$\mu$m are caused 
by variations in the disk structure, and thus a model can be constructed assuming the disk illuminated by the 
non-extincted stellar photosphere.  

A very simple inner disk model (not taking the outer disk into account) can be constructed
to estimate a lower-limit to the disk mass. Assuming that the innermost disk is optically thin in the IR and dominated
by a single temperature, the flux observed at a given frequency $\nu$, F$_\nu$, can be written as
\begin{equation}
F_\nu = \Omega B_\nu(T_{ring}) k_\nu \Sigma,
\end{equation}
where $\Omega$ is the solid angle subtended by the structure, $B_\nu(T_{ring})$ is the black body function evaluated
for the dominant ring temperature $T_{ring}$, $k_\nu$ is the dust mass absorption coefficient (cm$^2$/g) and $\Sigma$ is the
surface density of the ring. Since $\Omega$ depends on the area of the disk (with an inclination factor included), this means 
that $F_\nu$ is proportional to the mass of the ring. Taking $k_\nu$ to be a power law of the frequency with exponent 
$\beta$=1.9 and k$_{70\mu m}$=118 cm$^2$g$^{-1}$ \citep[for small interstellar grains without ice mantles;][]{roccatagliata13}, we 
find that the mid-IR emission can be relatively well 
fitted with a dominant ring temperature of 850K and a dust mass of 3e-3 M$_{Ceres}$,
or 4e-4 M$_{Ceres}$ for a higher temperature of 1270 K. 
Considering the corotation temperature, 1540 K, the amount of dust required would decrease to
2e-4 M$_{Ceres}$. If the gas-to-dust ratio takes the usual value of 100, this would mean a mass
between  2-30\% M$_{Ceres}$  including gas and dust.
The result is strongly dependent on 
the ring temperature, and because a single temperature cannot fit equally well all the observed datapoints, we
expect the inner disk to cover a range of temperatures.
Further variations in the dust properties (highly
processed silicates or large grains will have a different emissivity),
and whether the disk is optically thin (the mass is a lower limit) will also play a role here. 
Nevertheless, this exercise 
reveals that, as for GW Ori \citep{fang14}, a small change in the dust content of the inner disk can be responsible for a
very remarkable change in the mid-IR fluxes. For the observed accretion rate, this amount of mass is 
comparable to what
can be accreted in few months to a few years, so piling up the extra material in $\sim$4 years time 
could be done with
a mismatch between disk accretion and stellar accretion happening at some place between the outer 
disk and the stellar
magnetosphere.

Since the assumptions of optically thin inner disk and constant temperature are  a poor approximation for a
relatively dense and likely extended protoplanetary disk, we also modeled the 
disk using the radiative transfer code RADMC-2D 
\citep{dullemond04,dullemond11}. Because the SED is so uncertain in terms of extinction and variability,
we just aim to estimate how much mass would be needed to reproduce the
variability in the innermost disk, and the misinclination of the disks is not taken into account. 
As a photospheric model, we use a MARCS photosphere \citep{gustafsson08} scaled to the observed luminosity.
We explored various dust distributions within the nearly-cleared inner hole of the
disk, and set up the scattered light ring as the maximum outer disk inner radius. The total mass needed depends
on the grain properties. We assume standard amorphous silicates and carbon (in a proportion 3:1) with
typical sizes between 0.1 $\mu$m to 1cm (following a collisional distribution with exponent -3.5), with opacities
derived from the Jena database \citep{jaeger03}. Adding a dust surface density of the order of 
1e-8 g/cm$^2$ at 0.06 au,
decreasing
with the disk radius as r$^{-2}$ or steeper (so that most of the mass is concentrated in the
 innermost few au), is enough to go from
photospheric fluxes to the observed ones, and the total mass involved would be of the order of 3-5$\times$10$^{-5}$ M$_{Ceres}$ if we
assume a gas-to-dust ratio of 100. Note that the RADMC model underestimates the near-IR fluxes, so an inner wall
(not included in the current model) and thus a larger mass are likely.
As a final point, although the far-IR part of the SED is highly uncertain due to the large beam and unknown variability,
these simple models also reveal that the 22-24$\mu$m flux is dominated by the outer disk inner wall, which 
explains the lack of variability between W4 and MIPS 24$\mu$m.

The model confirms that the innermost disk structural changes required to alter the IR SED appearance
as observed could be caused by small accretion
variations between the outer and the inner disk, without needing any  more
dramatic processes.  Although the SED can be fitted with an inner disk that is clearly devoid of matter compared
to the outer disk, there has to be enough material close to the star to support the observed accretion, 
although the present data does not allow us to set further constraints.

From this exercise, despite all uncertainties involved in the analysis of the 
non-simultaneous SED of a highly variable object, the conclusion we reach is that the variability
observed in the innermost disk is consistent with matter transport at rates similar to the observed 
accretion rate onto the star.
The timescales of the variations ($\sim$4 years) imply that the material cannot be 
located much further away than a couple of 
au from the star.  The mass in the inner
disk would be in any case many orders of magnitude lower than the disk mass, which 
is estimated to be 0.018 M$_\odot$ in the
RADMC models to account for the 880$\mu$m flux. The time constraint imposed by the 
observed variability would be consistent with the orbital time of an undetected companion
at a couple of au. Dusty material moving at few-au should definitely emit in the mid-IR, so that future time-resolved mid-IR data over several years may help to explore this scenario.

\section{Discussion \label{discussion}}

We now use all the previous information to trace a complete picture of
the RX~J1604.3-2130A system and to investigate
the physical mechanism(s) behind the variability observed.  
The lightcurve presents dramatic changes on very short timescales,
including phase shifts, lightcurve profile variations from period to period, and sort-lived period drifts,
even if the system eventually reverts to the 5d period. Such changes are too rapid to 
be explained by  long-lived, cold stellar spots. Moreover, the dusty composition of the occulting
material requires the structures to be located beyond the dust destruction radius, for instance, in a clumpy
or warped disk. The eclipsing material would
be located at the inner disk at the corotation radius, where star-disk interactions are expected
to be highly dynamical \citep{fonseca14, mcginnis15}, although current stellar parameters 
can be only reconciled with a rotational period if the radius is increased by $\sim$10-15\%.
The observations are consistent with one major and one minor warps on a dusty disk that is
highly inclined with respect to the outer disk.  The warps may be associated to quasi-stable accretion columns
\citep[as in][]{alencar12, bodman17,alencar18}.
This highly inclined, wobbly/warped disk can also explain the shadows observed in the outer disk \citep{pinilla18}. 
The dust observations are in agreement with ALMA gas observations suggesting that the innermost
gaseous disk is also misaligned with respect to the outer disk \citep{mayama18}.

Magneto hydro-dynamic (MHD) models usually distinguish between two accretion scenarios: stable (with accretion columns
that are fixed to the star for at least several rotational periods) and unstable \citep[where
accretion proceeds through several rapidly-changing fingers due to e.g. Rayleigh-Taylor instability between the
innermost disk and the star, see][]{kurosawa13}. Unstable accretion can
explain quick changes from one rotational period to the next. Nevertheless,
although we observe a drift in phase along the K2 observations, the
timing of the eclipses is not chaotic (Figure \ref{K2wrapped-fig}) but rather consistent with 
two structures 
on either side of the star. These structures do change from one rotation period to the next, but they are more
stable than Rayleigh-Taylor fingers distributed over the stellar surface.
The well-defined redshifted absorption features in some of the
H$\alpha$ and H$\beta$ spectra are also characteristic of a system with stable accretion columns viewed 
close to along the line-of-sight. 
Thus one possibility would be RX~J1604.3-2130A having intermediate characteristics between
stable and unstable accretion, such as relatively stable accretion
structures attached to a particular longitude and latitude
on the stellar surface (e.g. due to a localized, dipolar magnetic field) but locally unstable \citep[e.g. due to Rayleigh-Taylor instability as proposed by][or any other localized instability]{kurosawa13}. 
If the accretion columns are relatively stable and locked to the disk at corotation, the most intense eclipses 
would be related to the rotational period. Unstable accretion along an otherwise-well-defined column could
produce changes from one rotational period to the next and changes in the vertical structure of
the inner disk.

The period of increased eclipsing activity (from approximately mission 
day 2085 to 2105) is consistent with the picture of
relatively unstable accretion through two well-defined structures. The periodicity 
is dominated by the 2.5d signature during this epoch (see Figures \ref{K2stackedGLSP-fig} and \ref{wavelet-fig}). 
This suggests that a change in the inner disk mass (deeper and more
frequent eclipses) results in 
accretion instability on the side of the star that is usually quiescent.  
Triggering accretion by
Rayleigh-Taylor instability does not need a change in the 
stellar magnetic field, but  could result from a change on the viscosity and/or 
disk accretion rate \citep{kurosawa13}. Additional 
material flowing inwards from the outer disk could increase 
the mass and density of the inner disk from time to time.
Some of the eclipses observed during this time are particularly deep, which would be also in agreement 
with an increased amount of matter in the inner disk.

Irregular feeding of the innermost disk could be a way to trigger thus both the 2.5d eclipses and to keep the star in an
unstable state where Rayleigh-Taylor instability changes the properties of the accretion columns on timescales
of days. It is likely that the accretion rate will be higher if the star is
accreting through both sides. Nevertheless, an
increase in accretion by a factor of few in a star that does not have a particularly high accretion rate (so
the accretion luminosity is small compared to the stellar luminosity) and that has complex extinction is hard to measure,
besides the two spots cannot be observed simultaneously in a star that is nearly equator-on.
Detailed time-resolved spectroscopy could help to disentangle rotational modulations from the effect of 
increased accretion \citep{sicilia15}. 

Most of the line profile variations observed with HIRES can be reproduced by rotational
modulations, although both the H$\alpha$ variability and the periods of increased eclipsing activity suggest that
the accretion rate is variable on longer timescales.
 Moreover, the times at which the star was observed to 
have photospheric mid-IR colors are also coincident with the times at which H$\alpha$ is weakest, suggesting lower 
accretion in agreement with the picture that the time-variability observed in the inner disk depends on the mass 
transport on a larger radial scale.  Future observations
of similar events are required for confirmation.

Massive bodies in the inner disk could be 
responsible for small variations in the matter flow that could increase
the pressure over the stellar magnetosphere and promote accretion along the two magnetically-active regions on both
sides of the star.  A highly asymmetric magnetic field distributed over the surface of the
star may be also taking part of the regulation so that accretion only proceeds through intense magnetic field regions 
where the star can efficiently lock into the innermost disk, not necessarily matching the rate of accretion through the disk and onto the star. If the corotation radius is located
at the distance where the Keplerian period is 5d, the magnetosphere will be quite large  ($\sim$9-10 R$_*$) compared to what
is usually assumed for young stars. This may be an additional signature of a strong magnetic field, maybe similar
to LkCa 15 \citep{donati19}.

Despite the rough agreement between the accretion rate needed to transport matter through the inner disk and the mass
required to produce the shadows and eclipses, an important problem remains: dealing with the angular momentum 
transport when matter from a face-on  outer disk is transferred to a nearly edge-on inner disk. Companions 
between the outer and inner disk may 
help in the process, but accretion of angular momentum from the outer disk would also tend to change the 
orientation of the very low-mass inner disk. A strong stellar magnetic field ensuring disk locking may force the material
in a particular orientation, but probably only very close-in. Therefore, the structure of the inner disk and of the
material in the cavity may be tridimensional, highly complex, and highly dynamic, including bridges and streamers
as it has been suggested for another dipper, AA Tau \citep{loomis17}. Near- and mid-IR observations
following the object during several years could help to constrain the structure and matter flow within the cavity.
Due to the location of the outer disk, a companion aligned with the outer 
disk could help to explain the formation history of the system, but it would not help with the 
change in angular momentum for material that is transported from outer to inner disk.
Nevertheless, massive companions ($>$2-3 M$_{J}$) at $>$22 au have been ruled out \citep{kraus08,canovas17},
which may be a problem to connect the outer with the inner disk. RX~J1604.3-2130B appears to be located too
far away from the disk \citep{koehler00} to have a significant effect, unless it has a highly eccentric orbit.
The Gaia DR2 parallax and proper motions for RX~J1604.3-2130A ($\varpi$=6.662$\pm$0.057 mas, -12.33$\pm$0.10 mas/yr
and -23.83$\pm$0.05 mas/yr) and RX~J1604.3-2130B ($\varpi$=6.79$\pm$0.10, -12.64$\pm$0.18 mas/yr 
and -24.73$\pm$0.09 mas/yr) are consistent with a common origin, but do not constrain the orbital
properties. A complex formation history, with two different protostellar collapse and 
accretion episodes (one originating the inner disk plus the star, and the
second producing the outer disk), could explain the formation.
Studying such formation scenarios, as well as the transfer of matter from the outer to the inner disk, 
would require  a better knowledge of the detailed disk structure and companions, in addition to
hydrodynamical simulations.

\section{Summary and conclusions \label{conclu}}

This study reveals the power of time-resolve data together with years of archival multi-wavelength data in the task of
disentangling the properties of very complex systems such as RX J1603.4-2130A. Our main results are summarized below:

\begin{itemize}
\item The observed eclipses have a quasi-periodicity of 5d,  consistent with the
rotational period of the star, and can be explained as extinction by
dusty structures, probably
warps at the points where the accretion columns leave the disk. 
\item The eclipsing dust is
located at the dust destruction radius in corotation with the star. The corotation radius
is of the order of  9-10 R$_*$, suggesting a very extended magnetosphere. 
\item The location of the warps suggests that the
star is accreting on a relatively stable regime, although occasional instabilities, maybe induced
by variations in the amount of matter in the inner disk, may trigger enhanced accretion through a
secondary spot, also resulting in more frequent eclipses.
\item The amount of material required to produce the optical eclipses (of the order of few percent of M$_{Ceres}$,  including gas and dust) 
is comparable to what is brought in by accretion in
few days to few weeks (depending on the inner disk structure and gas to dust ratio, 
and on the variable accretion rate), which explains the rapid variability.
Leaving aside the uncertainties in dust properties, structure shape, and inclination, the 
amount of material needed to produce the  IR shadows is also roughly consistent
with the mass needed to reproduce the
optical eclipses and rapid variability. 
\item As observed for GW Ori \citep{fang14}, the dusty (and likely gaseous) contents of the inner disk are variable in
timescales of years. These long-term variations could be produced by small 
mismatches of the accretion rate between the outer and the inner
disk, and can explain the lack of eclipsing activity that coincides with the lack of near-IR excess observed in the 
past. Future simultaneous optical and mid-IR data would be required to confirm the transport 
of material throughout the inner disk.  In particular, simultaneous time-resolved photometry 
and spectroscopy may help us to pin down potential phase shifts between
H$\alpha$ (close to the star) and the eclipses (close to the dust destruction radius) that can reveal the
innermost disk and magnetosphere structure. Longer-term followup (years) in the optical and IR could be used to
understand transport from the outer to the inner disk.
\item Together, the time-resolved photometry and spectroscopy also confirm that the star is not 
aligned with the outer disk, but highly inclined and rather aligned with the low-mass, highly variable, inner disk.
 Systems with highly inclined disks with respect to the stellar rotation
also pose a problem to the standard picture of protostellar collapse and disk formation mechanisms, so a deeper understanding of them may help us to understand the angular momentum transfer and the effect of initial conditions at protostar or cluster levels on the formation and subsequent evolution of protoplanetary disks.

\end{itemize}

\begin{acknowledgements}
We thank the referee for the comments and suggestions that helped to clarify this paper,
and S. Matsumura and M.H. Kristiansen for useful comments and discussion.
ASA is partly supported by the STFC grant number ST/S000399/1 (The Planet-Disk Connection: Accretion, Disk Structure, and Planet Formation). CFM acknowledges an ESO fellowship and has received funding from the European Union’s Horizon 2020 research and innovation program under the Marie Sklodowska-Curie grant agreement No 823823 (RISE DUSTBUSTERS). CFM  was partly supported by the Deutsche Forschungs-Gemeinschaft (DFG, German Research Foundation) - Ref no. FOR 2634/1 TE 1024/1-1. MB  acknowledges funding from ANR of France under contract number ANR-16-CE31-0013 (Planet Forming disks). PP acknowledges support provided by the Alexander von Humboldt Foundation in the framework of the Sofja Kovalevskaja Award endowed by the Federal Ministry of Education and Research. JdB acknowledges funding from the European Research Council under ERC Starting Grant agreement 678194 (FALCONER). JB acknowledges funding from the European Research Council (ERC) under the European Union’s Horizon 2020 research and innovation programme (grant agreement No 742095; {\it SPIDI}: Star-Planets-Inner Disk-Interactions.\\
This paper includes data collected by the Kepler mission. Funding for the Kepler mission is 
provided by the NASA Science Mission directorate. 
This publication makes use of data products from the Two Micron All Sky Survey, which is a joint project of the University of Massachusetts and the Infrared Processing and Analysis Center/California Institute of Technology, funded by the National Aeronautics and Space Administration and the National Science Foundation. 
This research has made use of the Keck Observatory Archive (KOA), which is operated by the W. M. Keck Observatory and the NASA Exoplanet Science Institute (NExScI), under contract with the National Aeronautics and Space Administration. 
This research has made use of the SIMBAD database, operated at CDS, Strasbourg, France \citep{wenger00}. This research has made use of the VizieR catalogue access tool, CDS, Strasbourg, France \citep{ochsenbein00}. This work
makes use of the CSS survey, which is funded by the National Aeronautics and Space
Administration under Grant No. NNG05GF22G issued through the Science
Mission Directorate Near-Earth Objects Observations Program. 
 The CRTS
survey is supported by the U.S.~National Science Foundation under
grants AST-0909182 and AST-1313422. 
This publication makes use of data products from the Wide-field Infrared Survey Explorer, which is a joint project of the University of California, Los Angeles, and the Jet Propulsion Laboratory/California Institute of Technology, funded by the National Aeronautics and Space Administration
This research made use of Astropy\footnote{http://www.astropy.org} a community-developed core Python package for Astronomy \citep{astropy13, astropy18}, and PyAstronomy\footnote{https://github.com/sczesla/PyAstronomy}. 
\end{acknowledgements}

\Online
\onecolumn

\begin{appendix}

\section{Data tables \label{data-app}}

This section contains the full REM observational data.

\begin{longtable}{ccccc} 
\caption{\label{griz-app-table} REM optical data. All magnitudes are given relative to
those of 58346.021046 (marked in the table with $^*$) for which g'=12.47$\pm$0.02 mag, r'=11.01$\pm$0.04 mag, i'=11.72$\pm$0.08: mag, and z'=11.01$\pm$0.08: mag. Note that g' is the only one for which the color terms are negligible, so that the
absolute calibration is highly uncertain for r'i'z'.  } \\                
\hline \hline                        
MJD & $\Delta$g' & $\Delta$r' & $\Delta$i' & $\Delta$z' \\
(d) & (mag) & (mag) & (mag) & (mag) \\
\hline                             
\endfirsthead
\caption{Continued.}\\
 \hline \hline                        
MJD & $\Delta$g' & $\Delta$r' & $\Delta$i' & $\Delta$z' \\
(d) & (mag) & (mag) & (mag) & (mag) \\
\hline                             
\endhead   
58247.213152 & 1.031$\pm$0.051 & 0.782$\pm$0.015 & 0.774$\pm$0.025 & 0.658$\pm$0.044  \\
58247.217176 & 1.076$\pm$0.038 & 0.794$\pm$0.066 & 0.790$\pm$0.025 & 0.813$\pm$0.065  \\
58249.213074 & 0.505$\pm$0.037 & 0.340$\pm$0.019 & 0.343$\pm$0.021 & 0.218$\pm$0.034  \\
58249.214534 & 0.457$\pm$0.023 & 0.328$\pm$0.016 & 0.398$\pm$0.018 & 0.361$\pm$0.035  \\
58249.215942 & 0.464$\pm$0.028 & 0.330$\pm$0.015 & 0.327$\pm$0.016 & 0.382$\pm$0.036  \\
58250.226890 & 0.184$\pm$0.019 & 0.080$\pm$0.010 & 0.168$\pm$0.011 & 0.145$\pm$0.016  \\
58250.228301 & 0.164$\pm$0.017 & 0.089$\pm$0.008 & 0.159$\pm$0.013 & 0.084$\pm$0.030  \\
58250.229714 & 0.091$\pm$0.017 & 0.077$\pm$0.013 & 0.165$\pm$0.013 & 0.182$\pm$0.034  \\
58251.278677 & 0.599$\pm$0.020 & 0.447$\pm$0.013 & 0.499$\pm$0.011 & 0.606$\pm$0.021  \\
58251.280120 & 0.579$\pm$0.024 & 0.500$\pm$0.024 & 0.536$\pm$0.012 & 0.681$\pm$0.039  \\
58251.281533 & 0.328$\pm$0.033 & 0.288$\pm$0.017 & 0.258$\pm$0.017 & 0.227$\pm$0.031  \\
58253.214545 & 0.151$\pm$0.024 & 0.080$\pm$0.011 & 0.191$\pm$0.018 & 0.104$\pm$0.017  \\
58253.215956 & 0.094$\pm$0.023 & 0.008$\pm$0.021 & 0.115$\pm$0.021 & 0.060$\pm$0.016  \\
58254.971876 & 1.055$\pm$0.026 & 0.485$\pm$0.015 & 0.369$\pm$0.018 & 0.189$\pm$0.018  \\
58254.974112 & 0.956$\pm$0.029 & 0.530$\pm$0.012 & 0.364$\pm$0.018 & 0.247$\pm$0.021  \\
58254.975229 & 0.939$\pm$0.031 & 0.490$\pm$0.010 & 0.297$\pm$0.015 & 0.182$\pm$0.017  \\
58254.976738 & 0.921$\pm$0.021 & 0.501$\pm$0.008 & 0.335$\pm$0.015 & 0.192$\pm$0.025  \\
58254.977855 & 0.867$\pm$0.016 & 0.420$\pm$0.011 & 0.292$\pm$0.013 & 0.218$\pm$0.020  \\
58254.978971 & 0.791$\pm$0.020 & 0.402$\pm$0.011 & 0.267$\pm$0.015 & 0.134$\pm$0.024  \\
58254.980089 & 0.844$\pm$0.018 & 0.433$\pm$0.012 & 0.277$\pm$0.015 & 0.098$\pm$0.024  \\
58254.981751 & 0.817$\pm$0.032 & 0.390$\pm$0.010 & 0.291$\pm$0.012 & 0.196$\pm$0.019  \\
58254.982867 & 0.757$\pm$0.022 & 0.338$\pm$0.014 & 0.290$\pm$0.009 & 0.203$\pm$0.025  \\
58254.983983 & 0.695$\pm$0.025 & 0.331$\pm$0.010 & 0.277$\pm$0.013 & 0.138$\pm$0.021  \\
58254.985100 & 0.708$\pm$0.026 & 0.339$\pm$0.010 & 0.300$\pm$0.011 & 0.235$\pm$0.027  \\
58254.986513 & 0.701$\pm$0.022 & 0.326$\pm$0.011 & 0.260$\pm$0.010 & 0.232$\pm$0.028  \\
58254.987629 & 0.702$\pm$0.028 & 0.298$\pm$0.010 & 0.238$\pm$0.009 & 0.148$\pm$0.033  \\
58254.988746 & 0.636$\pm$0.017 & 0.272$\pm$0.008 & 0.225$\pm$0.012 & 0.246$\pm$0.018  \\
58254.989866 & 0.587$\pm$0.017 & 0.265$\pm$0.010 & 0.240$\pm$0.011 & 0.151$\pm$0.025  \\
58254.991300 & 0.572$\pm$0.022 & 0.259$\pm$0.008 & 0.226$\pm$0.019 & 0.179$\pm$0.017  \\
58254.992419 & 0.567$\pm$0.027 & 0.215$\pm$0.011 & 0.210$\pm$0.020 & 0.179$\pm$0.031  \\
58254.993537 & 0.593$\pm$0.025 & 0.236$\pm$0.009 & 0.200$\pm$0.012 & 0.224$\pm$0.022  \\
58254.994654 & 0.548$\pm$0.015 & 0.201$\pm$0.008 & 0.185$\pm$0.013 & 0.226$\pm$0.018  \\
58254.996089 & 0.514$\pm$0.019 & 0.240$\pm$0.010 & 0.230$\pm$0.016 & 0.152$\pm$0.024  \\
58254.997206 & 0.532$\pm$0.019 & 0.179$\pm$0.007 & 0.169$\pm$0.010 & 0.096$\pm$0.023  \\
58254.998324 & 0.461$\pm$0.026 & 0.181$\pm$0.008 & 0.155$\pm$0.012 & 0.137$\pm$0.019  \\
58254.999440 & 0.488$\pm$0.024 & 0.183$\pm$0.009 & 0.169$\pm$0.013 & 0.127$\pm$0.025  \\
58255.000876 & 0.510$\pm$0.020 & 0.175$\pm$0.009 & 0.160$\pm$0.009 & 0.138$\pm$0.019  \\
58255.001993 & 0.504$\pm$0.021 & 0.171$\pm$0.010 & 0.155$\pm$0.017 & 0.130$\pm$0.018  \\
58255.003114 & 0.463$\pm$0.017 & 0.130$\pm$0.010 & 0.151$\pm$0.010 & 0.149$\pm$0.019  \\
58255.004232 & 0.450$\pm$0.024 & 0.138$\pm$0.011 & 0.173$\pm$0.008 & 0.198$\pm$0.019  \\
58255.005641 & 0.439$\pm$0.018 & 0.135$\pm$0.009 & 0.174$\pm$0.009 & 0.073$\pm$0.023  \\
58255.006758 & 0.397$\pm$0.023 & 0.143$\pm$0.009 & 0.120$\pm$0.010 & 0.093$\pm$0.021  \\
58255.007876 & 0.389$\pm$0.021 & 0.117$\pm$0.009 & 0.166$\pm$0.012 & 0.153$\pm$0.015  \\
58255.008993 & 0.336$\pm$0.018 & 0.108$\pm$0.009 & 0.098$\pm$0.012 & 0.103$\pm$0.017  \\
58255.011542 & 0.386$\pm$0.021 & 0.072$\pm$0.008 & 0.120$\pm$0.013 & 0.120$\pm$0.017  \\
58255.012660 & 0.394$\pm$0.017 & 0.081$\pm$0.009 & 0.102$\pm$0.015 & 0.156$\pm$0.017  \\
58255.013780 & 0.362$\pm$0.019 & 0.102$\pm$0.008 & 0.139$\pm$0.009 & 0.073$\pm$0.016  \\
58255.015190 & 0.246$\pm$0.021 & 0.044$\pm$0.012 & 0.114$\pm$0.013 & 0.088$\pm$0.026  \\
58255.016306 & 0.296$\pm$0.018 & 0.060$\pm$0.008 & 0.113$\pm$0.011 & 0.078$\pm$0.015  \\
58255.017428 & 0.296$\pm$0.019 & 0.060$\pm$0.007 & 0.108$\pm$0.008 & 0.046$\pm$0.019  \\
58255.018546 & 0.258$\pm$0.020 & 0.032$\pm$0.008 & 0.102$\pm$0.010 & 0.030$\pm$0.023  \\
58255.019980 & 0.258$\pm$0.017 & 0.062$\pm$0.007 & 0.084$\pm$0.009 & 0.026$\pm$0.017  \\
58255.021096 & 0.242$\pm$0.020 & 0.000$\pm$0.009 & 0.049$\pm$0.011 & 0.079$\pm$0.019  \\
58255.022214 & 0.190$\pm$0.017 & 0.029$\pm$0.007 & 0.086$\pm$0.010 & 0.057$\pm$0.020  \\
58255.023331 & 0.248$\pm$0.019 & 0.027$\pm$0.010 & 0.068$\pm$0.010 & 0.077$\pm$0.018  \\
58255.024760 & 0.209$\pm$0.015 & 0.010$\pm$0.007 & 0.103$\pm$0.011 & 0.106$\pm$0.019  \\
58255.025879 & 0.199$\pm$0.016 & 0.017$\pm$0.008 & 0.072$\pm$0.016 & 0.092$\pm$0.018  \\
58255.026997 & 0.221$\pm$0.017 & -0.007$\pm$0.006 & 0.070$\pm$0.010 & 0.032$\pm$0.017  \\
58255.028114 & 0.265$\pm$0.020 & -0.008$\pm$0.009 & 0.080$\pm$0.008 & 0.079$\pm$0.023  \\
58255.029525 & 0.187$\pm$0.017 & -0.024$\pm$0.009 & 0.048$\pm$0.009 & 0.056$\pm$0.014  \\
58255.030648 & 0.210$\pm$0.016 & 0.022$\pm$0.009 & 0.089$\pm$0.012 & 0.053$\pm$0.022  \\
58255.031773 & 0.191$\pm$0.023 & 0.014$\pm$0.008 & 0.087$\pm$0.007 & 0.044$\pm$0.015  \\
58255.032889 & 0.186$\pm$0.017 & 0.016$\pm$0.010 & 0.089$\pm$0.012 & 0.096$\pm$0.016  \\
58255.034298 & 0.194$\pm$0.018 & -0.002$\pm$0.008 & 0.104$\pm$0.014 & 0.018$\pm$0.017  \\
58255.035414 & 0.134$\pm$0.020 & -0.036$\pm$0.009 & 0.048$\pm$0.014 & -0.005$\pm$0.019  \\
58255.036532 & 0.163$\pm$0.017 & -0.002$\pm$0.007 & 0.055$\pm$0.007 & 0.057$\pm$0.017  \\
58255.037653 & 0.179$\pm$0.012 & -0.007$\pm$0.011 & 0.031$\pm$0.008 & 0.009$\pm$0.017  \\
58255.039090 & 0.136$\pm$0.017 & -0.030$\pm$0.009 & 0.053$\pm$0.009 & -0.023$\pm$0.016  \\
58255.040207 & 0.168$\pm$0.018 & -0.026$\pm$0.007 & 0.053$\pm$0.009 & 0.004$\pm$0.022  \\
58255.041325 & 0.174$\pm$0.015 & -0.038$\pm$0.010 & 0.050$\pm$0.009 & -0.014$\pm$0.022  \\
58255.042445 & 0.117$\pm$0.013 & -0.061$\pm$0.009 & 0.063$\pm$0.008 & -0.029$\pm$0.015  \\
58255.043857 & 0.126$\pm$0.015 & -0.051$\pm$0.008 & 0.043$\pm$0.010 & -0.042$\pm$0.015  \\
58255.044974 & 0.115$\pm$0.014 & -0.041$\pm$0.008 & 0.041$\pm$0.011 & -0.036$\pm$0.016  \\
58255.046095 & 0.138$\pm$0.017 & -0.054$\pm$0.011 & 0.050$\pm$0.010 & -0.072$\pm$0.021  \\
58255.047221 & 0.100$\pm$0.015 & -0.055$\pm$0.009 & 0.024$\pm$0.008 & -0.068$\pm$0.019  \\
58255.048632 & 0.105$\pm$0.033 & -0.079$\pm$0.012 & 0.015$\pm$0.009 & -0.025$\pm$0.022  \\
58255.049753 & 0.075$\pm$0.019 & -0.080$\pm$0.009 & 0.035$\pm$0.011 & -0.039$\pm$0.020  \\
58255.050870 & 0.078$\pm$0.016 & -0.077$\pm$0.010 & 0.012$\pm$0.009 & -0.107$\pm$0.019  \\
58255.051987 & 0.064$\pm$0.015 & -0.068$\pm$0.009 & 0.003$\pm$0.009 & -0.049$\pm$0.014  \\
58255.053401 & -0.122$\pm$0.023 & -0.103$\pm$0.010 & 0.032$\pm$0.010 & -0.070$\pm$0.014  \\
58255.054519 & 0.041$\pm$0.012 & -0.063$\pm$0.010 & 0.016$\pm$0.008 & -0.061$\pm$0.015  \\
58255.055635 & 0.030$\pm$0.017 & -0.088$\pm$0.008 & -0.008$\pm$0.017 & -0.044$\pm$0.014  \\
58255.056752 & 0.050$\pm$0.015 & -0.084$\pm$0.010 & 0.023$\pm$0.010 & -0.021$\pm$0.017  \\
58255.058162 & 0.056$\pm$0.014 & -0.076$\pm$0.011 & 0.021$\pm$0.008 & -0.042$\pm$0.016  \\
58255.059282 & 0.048$\pm$0.020 & -0.081$\pm$0.010 & 0.037$\pm$0.015 & -0.045$\pm$0.015  \\
58255.060398 & 0.028$\pm$0.014 & -0.094$\pm$0.011 & 0.011$\pm$0.009 & -0.075$\pm$0.013  \\
58255.061516 & 0.032$\pm$0.013 & -0.076$\pm$0.008 & 0.011$\pm$0.008 & -0.104$\pm$0.015  \\
58255.062924 & 0.022$\pm$0.016 & -0.087$\pm$0.010 & -0.002$\pm$0.008 & -0.039$\pm$0.013  \\
58255.064041 & 0.008$\pm$0.017 & -0.110$\pm$0.010 & -0.011$\pm$0.009 & -0.052$\pm$0.018  \\
58255.065158 & 0.019$\pm$0.014 & -0.129$\pm$0.013 & -0.009$\pm$0.011 & -0.097$\pm$0.015  \\
58255.066277 & -0.018$\pm$0.020 & -0.097$\pm$0.009 & -0.041$\pm$0.007 & -0.055$\pm$0.014  \\
58255.067685 & 0.008$\pm$0.017 & -0.106$\pm$0.008 & 0.019$\pm$0.010 & -0.071$\pm$0.019  \\
58255.068803 & 0.000$\pm$0.013 & -0.104$\pm$0.010 & -0.002$\pm$0.012 & -0.089$\pm$0.017  \\
58255.069920 & 0.004$\pm$0.013 & -0.093$\pm$0.008 & 0.016$\pm$0.018 & -0.078$\pm$0.016  \\
58255.071039 & -0.021$\pm$0.015 & -0.130$\pm$0.009 & -0.018$\pm$0.009 & -0.082$\pm$0.022  \\
58255.072482 & -0.023$\pm$0.016 & -0.090$\pm$0.009 & 0.010$\pm$0.008 & -0.021$\pm$0.014  \\
58255.073599 & -0.030$\pm$0.012 & -0.119$\pm$0.009 & -0.009$\pm$0.010 & -0.042$\pm$0.014  \\
58255.074718 & 0.038$\pm$0.014 & -0.120$\pm$0.008 & -0.014$\pm$0.011 & -0.046$\pm$0.016  \\
58255.075838 & -0.021$\pm$0.015 & -0.130$\pm$0.010 & 0.028$\pm$0.014 & 0.029$\pm$0.018  \\
58255.077269 & 0.003$\pm$0.016 & -0.107$\pm$0.011 & 0.036$\pm$0.017 & -0.090$\pm$0.019  \\
58255.078388 & -0.006$\pm$0.013 & -0.138$\pm$0.010 & -0.006$\pm$0.051 & -0.054$\pm$0.023  \\
58255.079504 & 0.018$\pm$0.014 & -0.126$\pm$0.009 & 0.015$\pm$0.016 & -0.006$\pm$0.017  \\
58255.080623 & -0.022$\pm$0.014 & -0.139$\pm$0.010 & -0.022$\pm$0.013 & -0.012$\pm$0.017  \\
58255.082033 & -0.037$\pm$0.015 & -0.133$\pm$0.011 & 0.008$\pm$0.013 & -0.089$\pm$0.016  \\
58255.083149 & -0.056$\pm$0.014 & -0.156$\pm$0.010 & 0.001$\pm$0.013 & -0.003$\pm$0.020  \\
58255.084266 & -0.057$\pm$0.014 & -0.143$\pm$0.012 & 0.008$\pm$0.015 & -0.022$\pm$0.024  \\
58255.085385 & -0.036$\pm$0.015 & -0.145$\pm$0.010 & -0.021$\pm$0.015 & -0.064$\pm$0.039  \\
58255.086793 & -0.049$\pm$0.014 & -0.142$\pm$0.010 & -0.049$\pm$0.015 & -0.057$\pm$0.012  \\
58255.087914 & -0.050$\pm$0.012 & -0.155$\pm$0.010 & -0.008$\pm$0.012 & -0.070$\pm$0.020  \\
58255.089031 & -0.070$\pm$0.014 & -0.155$\pm$0.010 & 0.009$\pm$0.015 & -0.068$\pm$0.102  \\
58255.090150 & -0.042$\pm$0.012 & -0.146$\pm$0.011 & -0.024$\pm$0.014 & -0.061$\pm$0.018  \\
58255.091557 & -0.071$\pm$0.011 & -0.179$\pm$0.011 & -0.049$\pm$0.011 & -0.090$\pm$0.017  \\
58255.092673 & -0.041$\pm$0.013 & -0.157$\pm$0.011 & -0.013$\pm$0.015 & -0.052$\pm$0.015  \\
58255.093793 & -0.070$\pm$0.014 & -0.183$\pm$0.011 & -0.049$\pm$0.012 & -0.047$\pm$0.037  \\
58255.094909 & -0.087$\pm$0.010 & -0.182$\pm$0.009 & -0.033$\pm$0.013 & -0.083$\pm$0.018  \\
58255.096342 & -0.076$\pm$0.014 & -0.167$\pm$0.010 & -0.072$\pm$0.012 & -0.016$\pm$0.018  \\
58255.097462 & -0.068$\pm$0.017 & -0.164$\pm$0.010 & -0.049$\pm$0.011 & 0.002$\pm$0.019  \\
58255.098579 & -0.059$\pm$0.012 & -0.140$\pm$0.010 & -0.038$\pm$0.012 & -0.074$\pm$0.030  \\
58255.099697 & -0.086$\pm$0.015 & -0.182$\pm$0.009 & -0.044$\pm$0.015 & -0.088$\pm$0.019  \\
58255.101130 & -0.064$\pm$0.015 & -0.167$\pm$0.011 & -0.001$\pm$0.013 & -0.048$\pm$0.014  \\
58255.102256 & -0.043$\pm$0.012 & -0.157$\pm$0.009 & -0.015$\pm$0.011 & -0.076$\pm$0.013  \\
58255.103372 & -0.106$\pm$0.012 & -0.159$\pm$0.012 & -0.020$\pm$0.011 & -0.055$\pm$0.023  \\
58255.104488 & -0.049$\pm$0.013 & -0.154$\pm$0.009 & -0.030$\pm$0.018 & -0.041$\pm$0.025  \\
58255.105897 & -0.077$\pm$0.011 & -0.173$\pm$0.010 & -0.018$\pm$0.012 & -0.078$\pm$0.015  \\
58255.107012 & -0.103$\pm$0.012 & -0.155$\pm$0.010 & -0.033$\pm$0.011 & -0.095$\pm$0.013  \\
58255.108131 & -0.078$\pm$0.009 & -0.169$\pm$0.009 & -0.027$\pm$0.009 & -0.066$\pm$0.017  \\
58255.109247 & -0.040$\pm$0.013 & -0.166$\pm$0.011 & -0.016$\pm$0.013 & -0.022$\pm$0.026  \\
58255.110681 & -0.086$\pm$0.010 & -0.162$\pm$0.010 & 0.000$\pm$0.021 & -0.046$\pm$0.013  \\
58255.110681 & -0.086$\pm$0.010 & -0.162$\pm$0.010 & 0.000$\pm$0.021 & -0.046$\pm$0.013  \\
58255.111799 & -0.074$\pm$0.013 & -0.170$\pm$0.010 & -0.017$\pm$0.014 & -0.062$\pm$0.013  \\
58255.112916 & -0.090$\pm$0.013 & -0.157$\pm$0.007 & -0.044$\pm$0.012 & -0.077$\pm$0.016  \\
58255.114032 & -0.086$\pm$0.015 & -0.153$\pm$0.010 & -0.017$\pm$0.012 & -0.011$\pm$0.022  \\
58255.115441 & -0.090$\pm$0.010 & -0.150$\pm$0.010 & -0.032$\pm$0.012 & 0.017$\pm$0.018  \\
58255.116558 & -0.048$\pm$0.012 & -0.146$\pm$0.011 & -0.013$\pm$0.018 & -0.051$\pm$0.022  \\
58255.117675 & -0.056$\pm$0.012 & -0.138$\pm$0.013 & -0.049$\pm$0.010 & -0.039$\pm$0.012  \\
58255.118791 & -0.074$\pm$0.012 & -0.163$\pm$0.011 & -0.027$\pm$0.011 & -0.044$\pm$0.014  \\
58255.120225 & -0.084$\pm$0.014 & -0.173$\pm$0.010 & -0.067$\pm$0.014 & -0.039$\pm$0.019  \\
58255.121343 & -0.096$\pm$0.011 & -0.179$\pm$0.009 & -0.032$\pm$0.011 & -0.038$\pm$0.024  \\
58255.122462 & -0.106$\pm$0.015 & -0.162$\pm$0.008 & 0.003$\pm$0.011 & -0.040$\pm$0.014  \\
58255.123582 & -0.082$\pm$0.016 & -0.161$\pm$0.008 & -0.036$\pm$0.012 & -0.092$\pm$0.017  \\
58255.125016 & -0.082$\pm$0.013 & -0.131$\pm$0.010 & 0.004$\pm$0.015 & -0.034$\pm$0.015  \\
58255.126135 & -0.095$\pm$0.010 & -0.166$\pm$0.008 & -0.017$\pm$0.013 & -0.032$\pm$0.016  \\
58255.126135 & -0.095$\pm$0.010 & -0.166$\pm$0.008 & -0.017$\pm$0.013 & -0.032$\pm$0.016  \\
58255.128367 & -0.090$\pm$0.013 & -0.173$\pm$0.009 & -0.010$\pm$0.010 & -0.035$\pm$0.014  \\
58255.129776 & -0.105$\pm$0.010 & -0.180$\pm$0.011 & -0.007$\pm$0.013 & -0.047$\pm$0.016  \\
58255.130893 & -0.107$\pm$0.013 & -0.149$\pm$0.009 & -0.041$\pm$0.011 & -0.040$\pm$0.015  \\
58255.132013 & -0.085$\pm$0.014 & -0.145$\pm$0.010 & -0.020$\pm$0.014 & 0.013$\pm$0.014  \\
58255.133130 & -0.080$\pm$0.010 & -0.148$\pm$0.010 & -0.015$\pm$0.014 & -0.038$\pm$0.023  \\
58255.134565 & -0.081$\pm$0.011 & -0.154$\pm$0.010 & -0.053$\pm$0.010 & -0.053$\pm$0.013  \\
58255.134565 & -0.081$\pm$0.011 & -0.154$\pm$0.010 & -0.053$\pm$0.010 & -0.053$\pm$0.013  \\
58255.135685 & -0.055$\pm$0.014 & -0.152$\pm$0.007 & 0.010$\pm$0.013 & -0.016$\pm$0.013  \\
58255.136801 & -0.097$\pm$0.010 & -0.151$\pm$0.010 & -0.010$\pm$0.009 & -0.064$\pm$0.013  \\
58255.139330 & -0.046$\pm$0.012 & -0.147$\pm$0.009 & -0.024$\pm$0.012 & -0.095$\pm$0.013  \\
58255.140447 & -0.083$\pm$0.014 & -0.132$\pm$0.010 & -0.013$\pm$0.009 & -0.026$\pm$0.013  \\
58255.141564 & -0.076$\pm$0.010 & -0.141$\pm$0.007 & -0.034$\pm$0.012 & -0.052$\pm$0.014  \\
58255.142683 & -0.076$\pm$0.013 & -0.145$\pm$0.009 & -0.010$\pm$0.010 & -0.046$\pm$0.013  \\
58255.144092 & -0.085$\pm$0.012 & -0.137$\pm$0.010 & -0.036$\pm$0.008 & -0.036$\pm$0.013  \\
58255.145210 & -0.085$\pm$0.012 & -0.134$\pm$0.009 & -0.020$\pm$0.009 & -0.048$\pm$0.013  \\
58255.146326 & -0.058$\pm$0.014 & -0.119$\pm$0.010 & -0.027$\pm$0.011 & -0.100$\pm$0.013  \\
58255.148852 & -0.050$\pm$0.012 & -0.117$\pm$0.009 & 0.001$\pm$0.013 & -0.076$\pm$0.013  \\
58255.149970 & -0.028$\pm$0.014 & -0.135$\pm$0.009 & 0.003$\pm$0.012 & -0.030$\pm$0.020  \\
58255.151087 & -0.060$\pm$0.012 & -0.115$\pm$0.007 & -0.023$\pm$0.010 & -0.086$\pm$0.013  \\
58255.152206 & -0.071$\pm$0.014 & -0.112$\pm$0.010 & -0.014$\pm$0.010 & -0.107$\pm$0.014  \\
58255.153642 & -0.072$\pm$0.012 & -0.108$\pm$0.010 & -0.004$\pm$0.013 & -0.058$\pm$0.014  \\
58255.153642 & -0.072$\pm$0.012 & -0.108$\pm$0.010 & -0.004$\pm$0.013 & -0.058$\pm$0.014  \\
58255.155882 & -0.072$\pm$0.011 & -0.124$\pm$0.010 & -0.033$\pm$0.010 & -0.084$\pm$0.016  \\
58255.156998 & -0.058$\pm$0.018 & -0.119$\pm$0.009 & 0.011$\pm$0.013 & -0.040$\pm$0.016  \\
58255.159553 & -0.042$\pm$0.013 & -0.112$\pm$0.007 & 0.009$\pm$0.010 & -0.028$\pm$0.018  \\
58255.160673 & -0.060$\pm$0.016 & -0.135$\pm$0.009 & -0.017$\pm$0.013 & 0.022$\pm$0.014  \\
58255.161790 & -0.040$\pm$0.014 & -0.115$\pm$0.009 & 0.008$\pm$0.009 & -0.045$\pm$0.014  \\
58255.163198 & -0.048$\pm$0.013 & -0.118$\pm$0.010 & 0.000$\pm$0.008 & 0.010$\pm$0.014  \\
58255.164317 & -0.074$\pm$0.013 & -0.108$\pm$0.008 & 0.002$\pm$0.009 & -0.014$\pm$0.013  \\
58255.166563 & -0.052$\pm$0.014 & -0.110$\pm$0.010 & 0.007$\pm$0.010 & -0.011$\pm$0.014  \\
58255.167997 & -0.035$\pm$0.014 & -0.104$\pm$0.008 & -0.025$\pm$0.012 & 0.000$\pm$0.013  \\
58255.169113 & -0.054$\pm$0.010 & -0.123$\pm$0.010 & 0.016$\pm$0.013 & -0.034$\pm$0.024  \\
58255.171349 & -0.038$\pm$0.015 & -0.116$\pm$0.009 & 0.008$\pm$0.009 & -0.014$\pm$0.015  \\
58255.172761 & 0.016$\pm$0.017 & -0.092$\pm$0.007 & 0.021$\pm$0.013 & 0.036$\pm$0.016  \\
58255.174994 & -0.085$\pm$0.011 & -0.091$\pm$0.009 & 0.011$\pm$0.010 & 0.006$\pm$0.015  \\
58255.176117 & -0.081$\pm$0.013 & -0.102$\pm$0.009 & 0.018$\pm$0.013 & 0.016$\pm$0.013  \\
58255.177524 & -0.070$\pm$0.014 & -0.117$\pm$0.009 & 0.011$\pm$0.011 & -0.031$\pm$0.015  \\
58255.178643 & -0.012$\pm$0.015 & -0.118$\pm$0.009 & 0.003$\pm$0.009 & 0.036$\pm$0.013  \\
58255.179761 & -0.061$\pm$0.011 & -0.083$\pm$0.010 & 0.039$\pm$0.011 & 0.007$\pm$0.013  \\
58255.179761 & -0.061$\pm$0.011 & -0.083$\pm$0.010 & 0.039$\pm$0.011 & 0.007$\pm$0.013  \\
58255.180879 & -0.068$\pm$0.013 & -0.111$\pm$0.008 & -0.013$\pm$0.011 & -0.014$\pm$0.014  \\
58255.182287 & 0.002$\pm$0.014 & -0.088$\pm$0.008 & -0.018$\pm$0.008 & -0.046$\pm$0.015  \\
58255.183405 & -0.075$\pm$0.014 & -0.116$\pm$0.010 & 0.018$\pm$0.011 & -0.063$\pm$0.013  \\
58255.184524 & -0.021$\pm$0.015 & -0.117$\pm$0.010 & 0.048$\pm$0.010 & 0.020$\pm$0.014  \\
58255.185645 & -0.048$\pm$0.015 & -0.092$\pm$0.010 & 0.009$\pm$0.013 & -0.021$\pm$0.014  \\
58260.214717 & 1.141$\pm$0.030 & 1.003$\pm$0.014 & 1.047$\pm$0.017 & 0.913$\pm$0.025  \\
58264.214959 & 0.241$\pm$0.033 & 0.117$\pm$0.018 & 0.146$\pm$0.017 & 0.067$\pm$0.027  \\
58264.216367 & 0.137$\pm$0.025 & 0.078$\pm$0.010 & 0.120$\pm$0.015 & 0.005$\pm$0.026  \\
58272.149714 & 0.027$\pm$0.026 & --- & 0.018$\pm$0.027 & -0.056$\pm$0.021  \\
58272.151123 & 0.284$\pm$0.024 & --- & 0.323$\pm$0.024 &  0.309$\pm$0.029  \\
58272.152533 & -0.113$\pm$0.025 & --- & -0.021$\pm$0.028 & -0.045$\pm$0.029 \\
58274.152662 & -0.144$\pm$0.019 & -0.156$\pm$0.010 & -0.069$\pm$0.023 & -0.272$\pm$0.015  \\
58274.155485 & -0.173$\pm$0.023 & -0.126$\pm$0.011 & -0.063$\pm$0.022 & -0.227$\pm$0.014  \\
58276.156968 & 0.365$\pm$0.027 & 0.253$\pm$0.018 & 0.230$\pm$0.014 & 0.101$\pm$0.043  \\
58276.158401 & 0.411$\pm$0.027 & 0.231$\pm$0.017 & 0.212$\pm$0.018 & 0.285$\pm$0.038  \\
58284.125002 & 0.091$\pm$0.024 & 0.082$\pm$0.015 & 0.105$\pm$0.016 & 0.140$\pm$0.029  \\
58284.126414 & 0.128$\pm$0.023 & 0.105$\pm$0.027 & 0.091$\pm$0.013 & 0.128$\pm$0.026  \\
58284.127833 & 0.146$\pm$0.023 & 0.065$\pm$0.027 & 0.058$\pm$0.009 & 0.122$\pm$0.035  \\
58286.130757 & 0.504$\pm$0.034 & 0.420$\pm$0.028 & 0.324$\pm$0.015 & 0.418$\pm$0.041  \\
58288.244378 & 0.204$\pm$0.029 & 0.137$\pm$0.009 & 0.180$\pm$0.018 & 0.122$\pm$0.024  \\
58288.245788 & 0.190$\pm$0.031 & 0.135$\pm$0.015 & 0.198$\pm$0.025 & 0.168$\pm$0.017  \\
58288.247220 & 0.161$\pm$0.029 & 0.111$\pm$0.012 & 0.133$\pm$0.023 & 0.103$\pm$0.022  \\
58290.267568 & 0.451$\pm$0.020 & 0.312$\pm$0.011 & 0.258$\pm$0.012 & 0.187$\pm$0.016  \\
58290.268978 & 0.517$\pm$0.021 & 0.404$\pm$0.013 & 0.323$\pm$0.018 & 0.194$\pm$0.019  \\
58292.296937 & 0.955$\pm$0.046 & 0.676$\pm$0.018 & 0.603$\pm$0.032 & 0.491$\pm$0.041  \\
58295.324367 & --- & 0.432$\pm$0.036 & --- & ---  \\
58298.086531 & --- & 0.453$\pm$0.022 & --- & ---  \\
58298.089355 & --- & 0.407$\pm$0.016 & --- & ---  \\
58307.240206 & 0.078$\pm$0.030 & -0.020$\pm$0.009 & 0.019$\pm$0.013 & -0.047$\pm$0.019  \\
58307.241641 & 0.109$\pm$0.023 & 0.008$\pm$0.012 & 0.004$\pm$0.015 & -0.052$\pm$0.039  \\
58307.243051 & 0.125$\pm$0.023 & 0.022$\pm$0.012 & 0.032$\pm$0.013 & -0.007$\pm$0.028  \\
58309.988264 & 0.371$\pm$0.025 & 0.164$\pm$0.018 & 0.207$\pm$0.009 & 0.238$\pm$0.025  \\
58309.989678 & 0.298$\pm$0.035 & 0.186$\pm$0.013 & 0.233$\pm$0.020 & 0.262$\pm$0.023  \\
58309.991112 & 0.324$\pm$0.036 & 0.174$\pm$0.021 & 0.298$\pm$0.009 & 0.332$\pm$0.029  \\
58311.995172 & -0.056$\pm$0.013 & -0.106$\pm$0.009 & 0.008$\pm$0.014 & 0.073$\pm$0.019  \\
58311.996584 & -0.108$\pm$0.018 & -0.132$\pm$0.009 & 0.013$\pm$0.018 & 0.000$\pm$0.021  \\
58311.998017 & -0.127$\pm$0.016 & -0.132$\pm$0.009 & 0.014$\pm$0.014 & -0.006$\pm$0.017  \\
58314.002261 & -0.041$\pm$0.014 & -0.017$\pm$0.009 & -0.045$\pm$0.010 & -0.111$\pm$0.028  \\
58314.002261 & -0.041$\pm$0.014 & -0.017$\pm$0.009 & -0.045$\pm$0.010 & -0.111$\pm$0.028  \\
58314.005102 & -0.027$\pm$0.013 & -0.013$\pm$0.007 & -0.054$\pm$0.014 & -0.075$\pm$0.033  \\
58321.001459 & -0.173$\pm$0.041 & -0.183$\pm$0.012 & -0.175$\pm$0.016 & -0.206$\pm$0.028  \\
58321.002866 & -0.228$\pm$0.043 & -0.180$\pm$0.013 & -0.156$\pm$0.022 & -0.118$\pm$0.029  \\
58323.007004 & --- & 0.086$\pm$0.022 & --- & ---  \\
58323.008441 & --- & 0.077$\pm$0.016 & --- & ---  \\
58323.009849 & --- & 0.081$\pm$0.015 & --- & ---  \\
58331.969675 & -0.019$\pm$0.039 & 0.027$\pm$0.016 & -0.011$\pm$0.022 & -0.188$\pm$0.023  \\
58336.225620 & 1.551$\pm$0.040 & 1.007$\pm$0.015 & 0.720$\pm$0.017 & 0.458$\pm$0.042  \\
58336.227028 & 1.665$\pm$0.028 & 1.057$\pm$0.014 & 0.809$\pm$0.018 & 0.380$\pm$0.037  \\
58338.989608 & 0.321$\pm$0.015 & 0.329$\pm$0.023 & 0.262$\pm$0.033 & 0.418$\pm$0.038  \\
58341.220383 & 0.990$\pm$0.023 & 0.578$\pm$0.009 & 0.366$\pm$0.013 & 0.210$\pm$0.019  \\
58341.223253 & 1.081$\pm$0.032 & 0.583$\pm$0.009 & 0.343$\pm$0.013 & 0.086$\pm$0.036  \\
58346.018229 & -0.035$\pm$0.011 & -0.018$\pm$0.007 & 0.010$\pm$0.007 & -0.025$\pm$0.012  \\
58346.019638 & 0.008$\pm$0.011 & -0.014$\pm$0.008 & 0.018$\pm$0.011 & -0.035$\pm$0.018  \\
58346.021046 & 0.000$\pm$0.009 & 0.000$\pm$0.005 & 0.000$\pm$0.005 & 0.000$\pm$0.012  \\
58348.048129 & 0.051$\pm$0.020 & 0.046$\pm$0.011 & 0.064$\pm$0.011 & 0.024$\pm$0.018  \\
58348.049538 & 0.020$\pm$0.019 & 0.033$\pm$0.013 & 0.016$\pm$0.011 & 0.030$\pm$0.019  \\
58348.049538 & 0.020$\pm$0.019 & 0.033$\pm$0.013 & 0.016$\pm$0.011 & 0.030$\pm$0.019  \\
58348.050946 & 0.027$\pm$0.019 & 0.013$\pm$0.011 & 0.064$\pm$0.010 & -0.024$\pm$0.016  \\
58352.085858 & --- & 1.306$\pm$0.014 & --- & ---  \\
58352.087294 & --- & 1.584$\pm$0.028 & --- & ---  \\
58354.088806 & 0.907$\pm$0.045 & 0.650$\pm$0.013 & 0.528$\pm$0.021 & 0.503$\pm$0.027  \\
58354.090241 & 0.849$\pm$0.021 & 0.696$\pm$0.016 & 0.569$\pm$0.016 & 0.449$\pm$0.025  \\
58356.092827 & 0.157$\pm$0.025 & 0.065$\pm$0.009 & 0.072$\pm$0.017 & 0.024$\pm$0.021  \\
58356.094261 & 0.145$\pm$0.018 & 0.062$\pm$0.008 & 0.044$\pm$0.019 & 0.018$\pm$0.027  \\
58356.095671 & 0.192$\pm$0.015 & 0.080$\pm$0.007 & 0.052$\pm$0.019 & 0.015$\pm$0.026  \\
58358.095785 & --- & 0.139$\pm$0.013 & --- & ---  \\
58358.097195 & --- & 0.128$\pm$0.016 & --- & ---  \\
58358.098627 & --- & 0.100$\pm$0.019 & --- & ---  \\
58360.149775 & 0.662$\pm$0.033 & 0.315$\pm$0.010 & 0.238$\pm$0.024 & 0.071$\pm$0.024  \\
58360.152597 & 0.698$\pm$0.024 & 0.359$\pm$0.015 & 0.315$\pm$0.017 & 0.183$\pm$0.027  \\
58363.042069 & 0.226$\pm$0.012 & 0.130$\pm$0.007 & 0.169$\pm$0.016 & 0.059$\pm$0.041  \\
58363.042069 & 0.226$\pm$0.012 & 0.130$\pm$0.007 & 0.169$\pm$0.016 & 0.059$\pm$0.041  \\
58367.136894 & 0.757$\pm$0.022 & 0.439$\pm$0.009 & 0.191$\pm$0.016 & 0.078$\pm$0.022  \\
58367.138303 & 0.776$\pm$0.021 & 0.438$\pm$0.009 & 0.196$\pm$0.020 & 0.094$\pm$0.021  \\
58367.139736 & 0.792$\pm$0.018 & 0.419$\pm$0.008 & 0.253$\pm$0.020 & 0.136$\pm$0.020  \\
58370.007730 & 0.892$\pm$0.048 & 0.732$\pm$0.021 & 0.732$\pm$0.033 & 0.550$\pm$0.063  \\
58370.009142 & 0.894$\pm$0.056 & 0.763$\pm$0.015 & 0.724$\pm$0.027 & 0.508$\pm$0.043  \\
58370.009142 & 0.894$\pm$0.056 & 0.763$\pm$0.015 & 0.724$\pm$0.027 & 0.509$\pm$0.043  \\
58373.115563 & 0.885$\pm$0.023 & 0.537$\pm$0.017 & 0.345$\pm$0.018 & 0.107$\pm$0.015  \\
58373.117020 & 0.943$\pm$0.025 & 0.549$\pm$0.018 & 0.351$\pm$0.023 & 0.080$\pm$0.021  \\
58378.074571 & 0.502$\pm$0.022 & 0.336$\pm$0.012 & 0.242$\pm$0.013 & 0.230$\pm$0.020  \\
58381.045817 & --- & 1.010$\pm$0.030 & --- & ---  \\
58383.090142 & 0.720$\pm$0.026 & 0.406$\pm$0.012 & 0.218$\pm$0.011 & 0.152$\pm$0.018  \\
58383.091578 & 0.773$\pm$0.033 & 0.419$\pm$0.013 & 0.220$\pm$0.016 & 0.002$\pm$0.024  \\
58383.092990 & 0.796$\pm$0.042 & 0.447$\pm$0.011 & 0.224$\pm$0.022 & 0.053$\pm$0.017  \\
58390.071665 & 1.373$\pm$0.026 & 0.996$\pm$0.028 & 0.799$\pm$0.019 & 0.657$\pm$0.018  \\
58392.074662 & 0.785$\pm$0.036 & 0.396$\pm$0.019 & 0.247$\pm$0.014 & -0.002$\pm$0.042  \\
58392.076100 & 0.839$\pm$0.016 & 0.384$\pm$0.013 & 0.271$\pm$0.025 & 0.050$\pm$0.020  \\

\hline                                             
\end{longtable}


\begin{longtable}{cccc} 
\caption{\label{JHK-app-table} REM near-IR data. Note that because the JHK exposures are not totally simultaneous, the MJD indicated is the one of the shortest wavelength observation available. All magnitudes are calibrated using 2MASS data, with calibration errors 2-4\% not included.} \\                
\hline \hline                        
MJD & J & H & K \\
(d) & (mag) & (mag) & (mag)  \\
\hline                             
\endfirsthead
\caption{Continued.}\\
 \hline \hline                        
MJD & J & H & K \\
(d) & (mag) & (mag) & (mag)  \\
\hline                             
\endhead   
58249.213255 & 9.132$\pm$0.025 & --- &  8.000$\pm$0.015  \\
58250.227253 & 8.920$\pm$0.020 & 8.205$\pm$0.007 & 7.895$\pm$0.012  \\
58253.213406 & 8.974$\pm$0.016 & 8.388$\pm$0.015 & 8.133$\pm$0.029  \\
58370.006668 & 9.350$\pm$0.020 & --- & ---  \\
58247.217478 & --- & 8.528$\pm$0.013 & ---  \\
58254.991590 & --- & 8.391$\pm$0.013 & ---  \\
58254.992182 & --- & 8.399$\pm$0.019 & ---  \\
58254.993962 & --- & 8.260$\pm$0.014 & ---  \\
58254.994560 & --- & 8.234$\pm$0.019 & ---  \\
58255.006540 & --- & 8.287$\pm$0.014 & ---  \\
58255.007138 & --- & 8.323$\pm$0.027 & ---  \\
58255.008326 & --- & 8.287$\pm$0.017 & ---  \\
58255.021456 & --- & 8.292$\pm$0.010 & ---  \\
58255.022048 & --- & 8.335$\pm$0.011 & ---  \\
58255.023234 & --- & 8.344$\pm$0.016 & ---  \\
58255.034590 & --- & 8.281$\pm$0.011 & ---  \\
58255.036384 & --- & 8.215$\pm$0.014 & ---  \\
58255.044760 & --- & 8.231$\pm$0.009 & ---  \\
58255.045948 & --- & 8.204$\pm$0.027 & ---  \\
58255.046552 & --- & 8.250$\pm$0.009 & ---  \\
58255.047156 & --- & 8.344$\pm$0.008 & ---  \\
58255.058474 & --- & 8.219$\pm$0.011 & ---  \\
58255.059066 & --- & 8.274$\pm$0.016 & ---  \\
58255.059658 & --- & 8.333$\pm$0.010 & ---  \\
58255.060250 & --- & 8.219$\pm$0.015 & ---  \\
58255.060852 & --- & 8.241$\pm$0.015 & ---  \\
58255.072776 & --- & 8.269$\pm$0.011 & ---  \\
58255.073370 & --- & 8.307$\pm$0.014 & ---  \\
58255.073962 & --- & 8.245$\pm$0.009 & ---  \\
58255.074554 & --- & 8.227$\pm$0.013 & ---  \\
58255.075144 & --- & 8.269$\pm$0.029 & ---  \\
58255.075754 & --- & 8.298$\pm$0.010 & ---  \\
58255.087110 & --- & 8.269$\pm$0.012 & ---  \\
58255.087708 & --- & 8.246$\pm$0.010 & ---  \\
58255.088302 & --- & 8.205$\pm$0.014 & ---  \\
58255.089492 & --- & 8.313$\pm$0.011 & ---  \\
58255.090094 & --- & 8.209$\pm$0.012 & ---  \\
58255.102022 & --- & 8.276$\pm$0.007 & ---  \\
58255.102622 & --- & 8.310$\pm$0.013 & ---  \\
58255.103214 & --- & 8.296$\pm$0.010 & ---  \\
58255.103812 & --- & 8.277$\pm$0.010 & ---  \\
58255.104404 & --- & 8.301$\pm$0.022 & ---  \\
58255.112152 & --- & 8.302$\pm$0.018 & ---  \\
58255.112746 & --- & 8.239$\pm$0.016 & ---  \\
58255.113356 & --- & 8.296$\pm$0.020 & ---  \\
58255.113948 & --- & 8.319$\pm$0.013 & ---  \\
58255.125308 & --- & 8.275$\pm$0.008 & ---  \\
58255.125898 & --- & 8.276$\pm$0.013 & ---  \\
58255.128288 & --- & 8.281$\pm$0.009 & ---  \\
58255.140244 & --- & 8.283$\pm$0.009 & ---  \\
58255.140836 & --- & 8.223$\pm$0.009 & ---  \\
58255.142030 & --- & 8.264$\pm$0.019 & ---  \\
58255.142630 & --- & 8.188$\pm$0.011 & ---  \\
58255.154528 & --- & 8.274$\pm$0.010 & ---  \\
58255.155126 & --- & 8.136$\pm$0.013 & ---  \\
58255.155722 & --- & 8.215$\pm$0.013 & ---  \\
58255.156314 & --- & 8.287$\pm$0.010 & ---  \\
58255.156922 & --- & 8.231$\pm$0.014 & ---  \\
58255.168290 & --- & 8.351$\pm$0.013 & ---  \\
58255.168910 & --- & 8.356$\pm$0.012 & ---  \\
58255.169476 & --- & 8.279$\pm$0.010 & ---  \\
58255.170682 & --- & 8.378$\pm$0.021 & ---  \\
58255.171280 & --- & 8.309$\pm$0.012 & ---  \\
58255.179642 & --- & 8.277$\pm$0.021 & ---  \\
58255.180848 & --- & 8.274$\pm$0.009 & ---  \\
58260.215018 & --- & 9.009$\pm$0.008 & ---  \\
58270.148478 & --- & 8.338$\pm$0.024 & ---  \\
58272.150000 & 9.365$\pm$0.014 & 8.256$\pm$0.041 & 8.152$\pm$0.026  \\
58276.155858 & 9.122$\pm$0.022 & 8.287$\pm$0.007 & 7.916$\pm$0.011  \\
58284.126700 & --- & 8.248$\pm$0.013 & ---  \\
58286.128202 & 9.145$\pm$0.018 & 8.135$\pm$0.023 & 7.938$\pm$0.024  \\
58288.244720 & 9.055$\pm$0.013 & 8.307$\pm$0.025 & 7.720$\pm$0.009  \\
58292.295794 & --- & 8.406$\pm$0.010 & ---  \\
58298.088246 & --- & 8.415$\pm$0.014 & ---  \\
58307.241924 & --- & 8.157$\pm$0.007 & ---  \\
58309.988535 & 9.042$\pm$0.028 & 8.438$\pm$0.009 & 7.812$\pm$0.047  \\
58314.003964 & --- & 8.351$\pm$0.008 & ---  \\
58325.060016 & --- & 8.597$\pm$0.008 & ---  \\
58331.971384 & --- & 8.092$\pm$0.043 & ---  \\
58350.071990 & --- & 8.307$\pm$0.013 & ---  \\
58358.097510 & --- & 8.361$\pm$0.015 & ---  \\
58375.119376 & --- & 8.514$\pm$0.020 & ---  \\
58375.119376 & --- & 8.512$\pm$0.020 & ---  \\
58381.044726 & --- & 9.034$\pm$0.014 & ---  \\
58392.076388 & --- & 8.572$\pm$0.017 & ---  \\
58251.281870 & --- & --- & 8.380$\pm$0.024  \\
58336.227364 & --- & --- & 8.133$\pm$0.034  \\
58363.042402 & 9.229$\pm$0.012 & 8.431$\pm$0.007 & 8.054$\pm$0.009  \\
58370.009478 & --- & --- & 8.084$\pm$0.041  \\
58373.117348 & --- & --- & 8.044$\pm$0.020  \\
58378.077774 & --- & --- & 8.155$\pm$0.037  \\
58383.093367 & --- & --- & 8.014$\pm$0.103  \\
\hline                                             
\end{longtable}


\section{Flares observed in the K2 data \label{flares-app}}

\begin{figure*}
\centering
\includegraphics[width=0.8\linewidth]{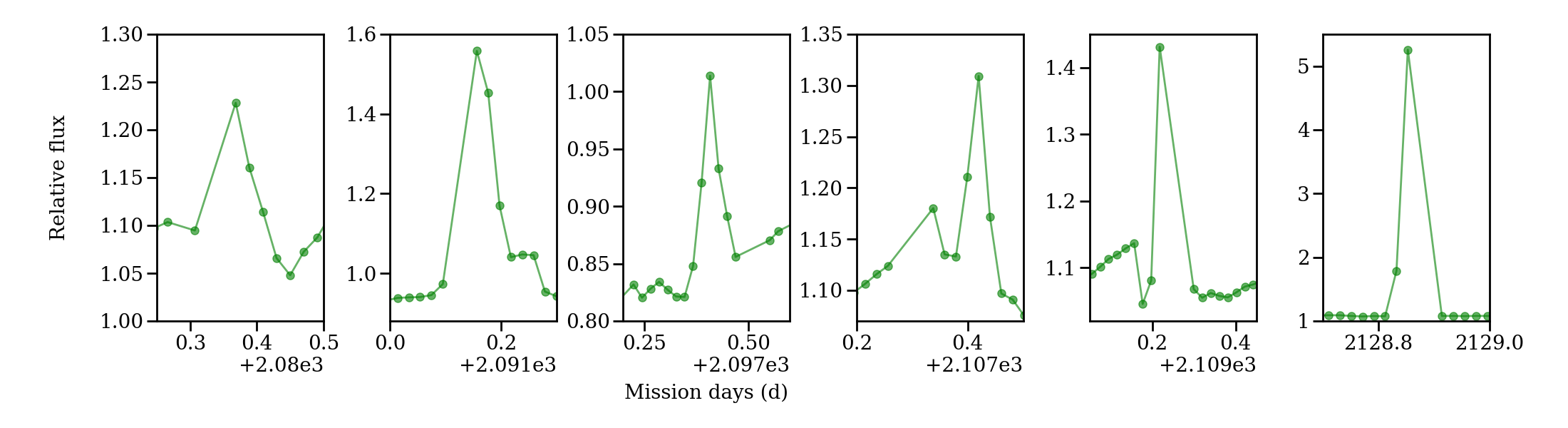}
\caption{Observed flares. \label{K2flares-fig}}
\end{figure*}

Despite the mild accretion, there are signatures of activity.
There are several brief increases in brightness in the K2 data that could be stellar flares (see Table \ref{flares-table}). 
All the cases were checked against contamination from Solar System objects, which can mimic the
flaring behavior \citep{rappaport19}, using SkyBot \citep{berthier06,berthier16} as done by \citep{szabo15},
and no contamination was detected (Martti Holst Kristiansen, private communication).
In 5 cases,
the increased flux is detected in at least 3 datapoints, revealing a profile consistent with stellar flares \citep{doyle18, gershberg83}, with a sharp rise and slower decline and duration between 1-3h (see
Figure \ref{K2flares-fig}). Four more potential events are detected in less than 3 observations \citep[consistent with short flares,][but also harder to confirm]{doyle18} and/or are weaker. One among them happens during an eclipse, so their classification is more uncertain.
One of them (on MJD 59021.611483) has extreme intensity, although it does not
show the usual flare profile (rapid rise, slower decay), 
but rises over two observations and falls of
from the peak within half an hour. The two largest 
flares happen at the beginning and at the end of the period between 2091-2010d when
the star also presents concatenated eclipses, but more data would be needed to tell whether this is significant. One of the flaring events may be composed of at least two consecutive flares, as it has been observed in some M-type stars \citep{doyle18}.  No flares are detected in the REM data, likely due to the sparse sampling.

\begin{table}
\caption{Flares detected in the K2 data. \label{flares-table}}
\centering
\begin{footnotesize}
\begin{tabular}{cccl}
\hline \hline 
MJD & Strength & Duration & Notes\\
(d) & (vs cont.) & (h) &  \\
\hline
58973.147562 & 1.2 & 1.5 & Weak\\
58983.935503 & 1.5 & 1.5 & Possible tail up to 1.5h more\\
58985.406582 & 1.5 & 1.5 & During eclipse, uncertain \\
58990.146722 & 1.2 & 2.5 & Rise and fall observed\\
59000.117352 & 1.2 & 2.5 & Messy, could be 2 flares\\
59001.997061 & 1.3 & 0.5: & Single datapoint, uncertain\\
59015.502389 & 1.4 & 0.5: & Single datapoint, uncertain\\
59021.611483 & 4.9 & 2.0 & Very strong, 2 points\\
59023.082572 & 1.1 & 1.5 & Weak, 2 points\\
\hline
\end{tabular}
\tablefoot{The MJD is given for the first point that
shows a significant increase in the flux. The strength is given
as the ratio with
respect to the local continuum. The length reflects the
time span until the flux is observed to be back to normal,
and is limited by the 30 min K2 cadence.}
\end{footnotesize}
\end{table}

\section{Summary of the kinematics from optical spectroscopy \label{kinematics-app}}

Table \ref{kinematics-table} lists the rotational and radial velocities derived for all the individual Keck spectra.
The rotational and radial velocities are consistent throughout all the spectra,
although the quality of the data from MJD 55287.614 is better than the others. There is no evidence of 
radial velocity variability within the observed errors. Constraining the possible radial velocity companions
with sparse-sampled data in an object with variable
accretion and quasi-stable accretion columns along the line-of-sight would require a detailed analysis to
distinguish accretion and gas absorption from the disk from those of potential companions \citep[e.g. see][]{mora02,sicilia15} that is beyond the scope of this paper.
Note that the spectra from 54689.304 and 55256.556 are noisier than the rest in the
regions used for the velocity estimates, and thus the results are highly uncertain.

\begin{table}
\caption{HIRES/Keck radial and rotational velocities. \label{kinematics-table}}
\centering
\begin{tabular}{ccc}
\hline \hline
MJD & Rad. Vel. & v$sin i$\\
(d) & (km/s) & (km/s)   \\
\hline
53902.275 & -6.6$\pm$0.6 & 16.7$\pm$0.4  \\
53902.278 & -6.8$\pm$0.4 & 16.6$\pm$0.7\\
54642.417 & -4.4$\pm$0.7 & 17.7$\pm$0.5\\
54689.304 & -5.7$\pm$1.5$^*$ & 18$\pm$2$^*$\\
55256.556 & -7.7$\pm$0.3$^*$ & 11$\pm$2$^*$\\
55287.614 & -6.8$\pm$0.1 & 16.2$\pm$0.6\\
\hline
\end{tabular}
\tablefoot{Values marked with $^*$ are uncertain due to poorer S/N in the spectra.}
\end{table}

\section{Archival SED photometry \label{SED-app}}

Table \ref{sed-table} contains the archival photometry from VizieR that we selected as being most likely
representative for the out-of-eclipse SED. We chose the 
brightest magnitudes observed in each filter, adding also those that are unlikely to change due to extinction
and on short timescales (mid-IR and longer wavelengths). Due to the non-homogeneity of the sample and to the
lack of detailed knowledge of the full spectrum variability, some of the selected points may have been taken
during eclipse phases.

\begin{table}
\caption{Archival SED data during the out-of-eclipse phases.\label{sed-table}}
\centering
\begin{footnotesize}
\begin{tabular}{llll}
\hline \hline 
$\lambda$ & F$_\nu$ & Survey & Reference \\
(\AA) & (Jy) & & \\
\hline
0.350 & 0.0014$\pm$0.0001 & SkyMapper u  &  W18  \\
0.387 & 0.0045$\pm$0.0004 & SkyMapper v  &  W18  \\
0.420 & 0.0286 & HIPPARCOS BT  &  U01  \\
0.444 & 0.0171$\pm$0.0086 & Johnson B  &  H15  \\
0.444 & 0.0188$\pm$0.0076 & Johnson B  &  H15  \\
0.468 & 0.0215$\pm$0.0068 & POSS-II J  &  L08  \\
0.482 & 0.026$\pm$0.012 & SDSS g$^\prime$  &  H15  \\
0.497 & 0.0403$\pm$0.0027 & SkyMapper g  & W18    \\
0.505 & 0.0329$\pm$0.0015 & GAIA DR2 Gbp  & D19   \\
0.554 & 0.045$\pm$0.019 & Johnson V  & H15   \\
0.604 & 0.067$\pm$0.005 & SkyMapper r  &  W18  \\
0.623 & 0.059$\pm$0.001 & GAIA DR2 G  & D19   \\
0.625 & 0.071$\pm$0.028 & SDSS r$^\prime$ &   H15 \\
0.763 & 0.105$\pm$0.044 & SDSS i$^\prime$ &  H15  \\
0.772 & 0.110$\pm$0.004 & GAIA DR2 Grp  &  D19  \\
0.784 & 0.123$\pm$0.039 & POSS-II i  & L08   \\
0.865 & 0.164 & PAN-STARRS z  &  C16  \\
3.35 & 0.293$\pm$0.006 & WISE W1  &  C12  \\
4.6 & 0.251$\pm$0.004 & WISE W2  &  C12  \\
11.6 & 0.0614$\pm$0.0008 & WISE W3  & C12   \\
22.1 & 0.152$\pm$0.003 & WISE W4  &  C12  \\
23.7 & 0.173$\pm$0.006 & Spitzer/MIPS 24  &  E18  \\
65 & 2.7 & AKARI N60  &  Y10  \\
90 & 3.15$\pm$0.25 & AKARI WIDE-S  &  Y10  \\
134 & 5.3$\pm$1.0 & AKARI WIDE-L  &  Y10  \\
160 & 2.62	 & AKARI N160  & Y10   \\
880 & 0.219$\pm$0.001 &  ALMA  & B16   \\
\hline
\end{tabular}
\tablefoot{References: B16 \citep{barenfeld16}; C12 \citep{cutri12}; C16 \citep{chambers16}; 
D19 \citep{damiani19}; E18 \citep{esplin18};  L08 \citep{lasker08}; U01 \citep{urban01}; 
Y10 \citep{yamamura10}; W18 \citep{wolf18}; Z04 \citep{zacharias04}. Note that some of the
datapoints are provided without uncertainties in the original papers.
}
\end{footnotesize}
\end{table}

\end{appendix}

\end{document}